\documentclass[letterpaper]{article} 
\usepackage{aaai2026}    
\usepackage{times}
\usepackage{helvet}
\usepackage{courier}
\usepackage{xcolor}
\usepackage{booktabs} 
\usepackage{multirow}
\usepackage[table]{xcolor}
\usepackage{amsmath,amssymb}
\usepackage[many]{tcolorbox}
\usepackage{amsmath,tabularx,booktabs}
\usepackage{enumitem}
\usepackage{times}  
\usepackage{helvet}  
\usepackage{courier}  
 \usepackage{amsmath} 
\usepackage[hyphens]{url}  
\usepackage{graphicx} 
\usepackage{natbib}  
\usepackage{caption} 
\usepackage{algorithm}
\usepackage{algorithmic}

\usepackage{newfloat}
\usepackage{listings}
\DeclareCaptionStyle{ruled}{labelfont=normalfont,labelsep=colon,strut=off} 
\floatstyle{ruled}
\newfloat{listing}{tb}{lst}{}
\floatname{listing}{Listing}
\title{TWICE: Modeling the Temporal Evolution of Personalized User Behavior via Event-Driven Agents}
\author {
    Bingrui Jin\textsuperscript{\rm 1,2},
    Kunyao Lan\textsuperscript{\rm 2},
    Baihan Li\textsuperscript{\rm 1,2},
    Mengyue Wu\textsuperscript{\rm 2}
}
\affiliations {
    \textsuperscript{\rm 1}SJTU Paris Elite Institute of Technology (SPEIT), Shanghai Jiao Tong University, Shanghai, China\\
    \textsuperscript{\rm 2}X‑LANCE Lab, School of Computer Science, Shanghai Jiao Tong University; MoE Key Lab of Artificial Intelligence, AI Institute, Shanghai Jiao Tong University, Shanghai, China\\
    \{suzy\_jin, lankunyao,lbh0612, mengyuewu\}@sjtu.edu.cn
}

\begin{document}

\maketitle

\begin{abstract}
User simulators are widely used for data generation, evaluation, and agent-based interaction, but existing approaches often model users as static personas or rely on generic historical context, making it difficult to capture how individual behavior evolves over time. To address this limitation, we propose TWICE, an LLM-based framework for temporally grounded personalized user simulation. TWICE combines structured user profiling, an event-driven memory module organized around life events and behavioral shifts, and a two-stage workflow separating event-grounded content planning from personalized style adaptation. This design enables the simulator to model not only what a user says, but also how past experiences shape later expression. We evaluate TWICE on a large-scale longitudinal Twitter dataset and introduce a comprehensive evaluation framework that jointly measures authenticity, consistency, and humanlikeness. Results show that TWICE consistently outperforms strong baselines, suggesting that event-centered memory is a promising mechanism for modeling the temporal evolution of personalized user behavior.
\end{abstract}

\section{Introduction}\label{sec:intro}

Modeling personalized human behavior over extended time horizons is a long-standing challenge in human-centered artificial intelligence. Emerging applications such as conversational assistants~\citep{Schatzmann2006Survey}, recommendation systems~\citep{Ricci2015RecommenderHandbook}, and social simulations~\citep{Lazer2009ComputationalSocialScience} increasingly require models that represent users not as static profiles but as individuals whose behaviors evolve over time. 
While existing LLM-based user simulators have made progress in modeling macro-level social dynamics~\citep{Yang2025OASIS} and static persona grounding~\citep{Wang2025USP,Hu2025PopAlign}, they do not explicitly model how individual expression evolves over time in response to past events. This limitation persists even in recent efforts on dynamic or evolving agents~\citep{Park2023GenerativeAgents, Li2023CAMEL, li2024evolvingagentsinteractivesimulation}.

Human behavior is not static; expression changes over time in response to life events, emotional shifts, and personal circumstances~\citep{Chancellor2020MethodsIP,Chen2024MappingLC}.

We argue that realistic user simulation should not be viewed as persona conditioning or historical replay. Instead, it requires jointly modeling three interrelated dimensions: \textit{what} users say (semantic content), \textit{why} they say it (event triggers), and \textit{how} prior experiences shape later expression over time (temporal conditioning). From this perspective, the core contribution of TWICE lies not in any single component, but in operationalizing temporally grounded user simulation as a unified behavioral modeling framework.

To address this, we propose \textbf{\textsc{TWICE}}
(\textbf{T}emporal-dynamics–aware \textbf{W}orkflow for
\textbf{I}ndividualized, \textbf{C}ontext- and \textbf{E}vent-driven user simulation). TWICE departs from simple historical dialogue memory by introducing an event-driven memory module capable of representing structured events and their temporal relationships.

\begin{figure*}[!t]
    \centering
    \includegraphics[width=0.93\textwidth, trim=2.7cm 1.3cm 2.7cm 1.3cm, clip]{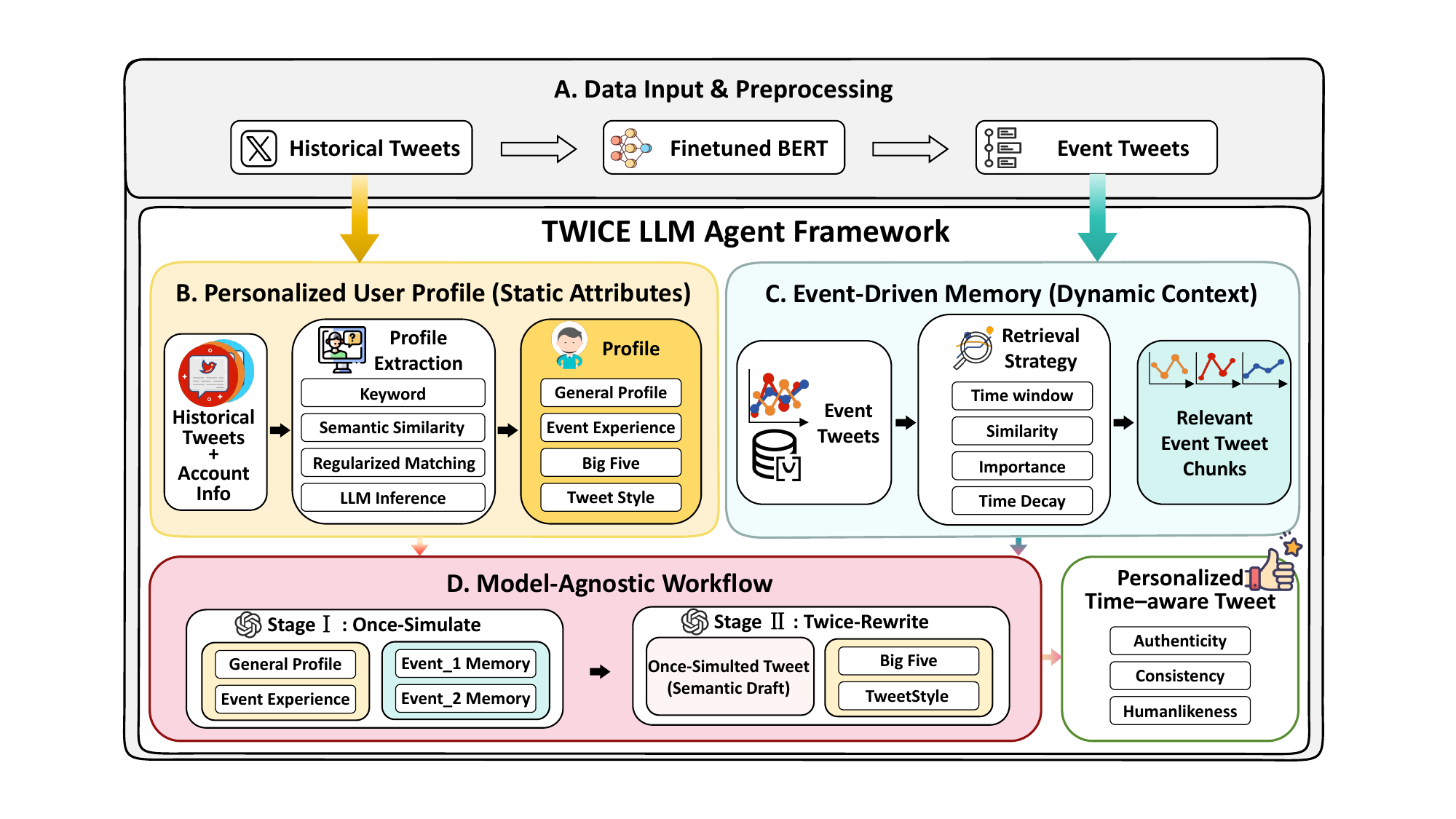}
    \caption{Overview of the \textbf{TWICE} Framework. This framework integrates personalized user profiling, an event-driven memory module, and a workflow for personalized style rewriting.}
    \label{fig:framework}
\end{figure*}

As illustrated in Figure~\ref{fig:framework}, our framework integrates three key components: (1) a structured user profile for stable user priors; (2) an event-driven memory module that balances semantic similarity with temporal locality; and (3) a two-stage workflow that decouples content planning from personalized style rewriting. By modeling social media posting as a temporally grounded process, we demonstrate that TWICE enables more authentic, consistent, and human-like user modeling across long horizons.


We evaluated TWICE on a large-scale Twitter dataset featuring 34,330 users associated with substantial temporal and affective variation in online expression. Our evaluation is guided by three research questions: (RQ1) how TWICE compares with existing personalized user simulation baselines in generating authentic and temporally grounded tweets; (RQ2) how event-driven memory contributes to temporal coherence and behavioral consistency; and (RQ3) whether TWICE generalizes across users with different levels of behavioral dynamics and emotional variation.

Importantly, our goal is not to model complete human cognition or personality, but to simulate temporally evolving patterns of observable online behavior grounded in user history and event context.


Results show that TWICE consistently outperforms strong baselines in generating temporally coherent and behaviorally consistent tweets. Our contributions are threefold: (1) we formalize long-term personalized tweet simulation as a temporally grounded behavioral modeling task; (2) we operationalize temporally grounded user simulation through an event-driven agent framework; and (3) we establish a multi-faceted evaluation protocol combining automatic metrics with human preference judgments.

\section{The TWICE Framework}\label{sec:method}
We formalize long-term user simulation as the task of modeling a user's evolving behavioral trajectory, where each generated post is conditioned on a stable personal identity and a dynamic set of temporal event triggers:
\begin{equation}
    y^{*}=\arg\max_{y}\;p\big(y \mid U,\,M_{\le t},\,E_t\big),
\end{equation}
where $y$ is a candidate tweet and $y^{*}$ is the optimal generated tweet; 
$U$ encodes stable user priors, $M_{\le t}$ summarizes temporally and semantically relevant history, and $E_t$ denotes the current event/context.

\subsection{Personalized User Profiling}

Accurate simulation requires stable user priors to mitigate style drift over months of activity. We construct a structured profile $U$ as a key-value store spanning both general and personalized attributes:
\begin{itemize}
    \item General Attributes: For demographics such as age, gender, and career, we utilize a pipeline of regular expressions for high recall, followed by embedding-based semantic matching to handle linguistic variability. Lightweight LLM inference is used for final disambiguation.
    \item Personalized Behavioral Attributes: To capture identity-level regularities, we incorporate Big Five personality traits, life-event histories, and stylistic evidence. We use supervised classifiers to detect high-confidence behavioral signals, which an LLM then summarizes into information-dense snippets.
    \item Stylistic Baseline: We iteratively select representative posts across a user's timeline and compress them into a natural-language style description accompanied by a set of stylistic exemplars.
\end{itemize}

Implementation details of feature extraction are provided in Appendix~\ref{app:profile}.

\subsection{Event-driven Memory Module}
Relying solely on recency or surface similarity often breaks the causal chains that shape human expression. Our memory module $M_{\le t}$ organizes history along two axes to ensure temporal coherence: 1) \textbf{General Memory:} Tweets are aggregated into fixed temporal windows (e.g., 30-day chunks) via pooled embeddings to capture baseline fluctuations; 2) \textbf{Event Memory:} Tweets are grouped by detected life events and behavioral shifts into unified nodes, preserving high-signal emotional trajectories.

\paragraph{Retrieval Strategy}

Given a current event/context $E_t$, we restrict retrieval to a backward temporal window $[t-\Delta, t)$ to ensure causal consistency. Memory nodes (general or event-level) are ranked by cosine similarity to $E_t$ and expanded into tweet-level candidates. Each candidate tweet is assigned a weighted score that combines semantic similarity, temporal decay, and importance:

\begin{equation}
\label{eq:score}
\begin{aligned}
\operatorname{score}(\text{tweet})
&= \cos\!\bigl(\mathbf{e}_{\text{tweet}}, \mathbf{e}_{E_t}\bigr)
\cdot e^{-\lambda \Delta t} \\
&\quad \cdot \bigl(1 + k(\mathrm{imp}-1)\bigr)
\cdot w_{\text{state}} .
\end{aligned}
\end{equation}

Here $\Delta t$ denotes the temporal gap between the candidate tweet and the current event, $\lambda$ controls the decay rate, $\mathrm{imp}$ is a per-tweet importance score derived from event relevance, $k$ scales the influence of importance weighting, and $w_{\text{state}}$ adjusts the score according to consistency with the inferred user state.

If the top-$K$ candidates are insufficient, we dynamically expand the search until $N$ memory entries are collected; the top $N$ form the final memory context $M_{\le t}$ for generation.

\subsection{Model-Agnostic Workflow}
To simulate how a user would express themselves in the present moment, we decouple factual planning from stylistic realization. This proceeds in two main stages:
\begin{itemize}
    \item \textbf{Stage I: Event-Grounded Content Generation.} Conditioned on $(U, M_{\le t}, E_t)$, the agent first produces a ``semantic draft''. This stage prioritizes factual alignment with the user's past and respects the emotional trajectory implied by retrieved memories, while intentionally avoiding specific idiolect to prevent overfitting. 
    \item \textbf{Stage II: Style Adaptation.} The semantic draft is rewritten into the user's distinct voice without altering its meaning. This is guided by three identity signals in $U$: Big Five personality cues for stance and intensity, the natural-language style description for controllable guidance, and stylistic exemplars to capture implicit habits like preferred pacing or formulaic phrases.
\end{itemize}

\begin{figure*}[t]
    \centering
    \includegraphics[width=0.90\textwidth]{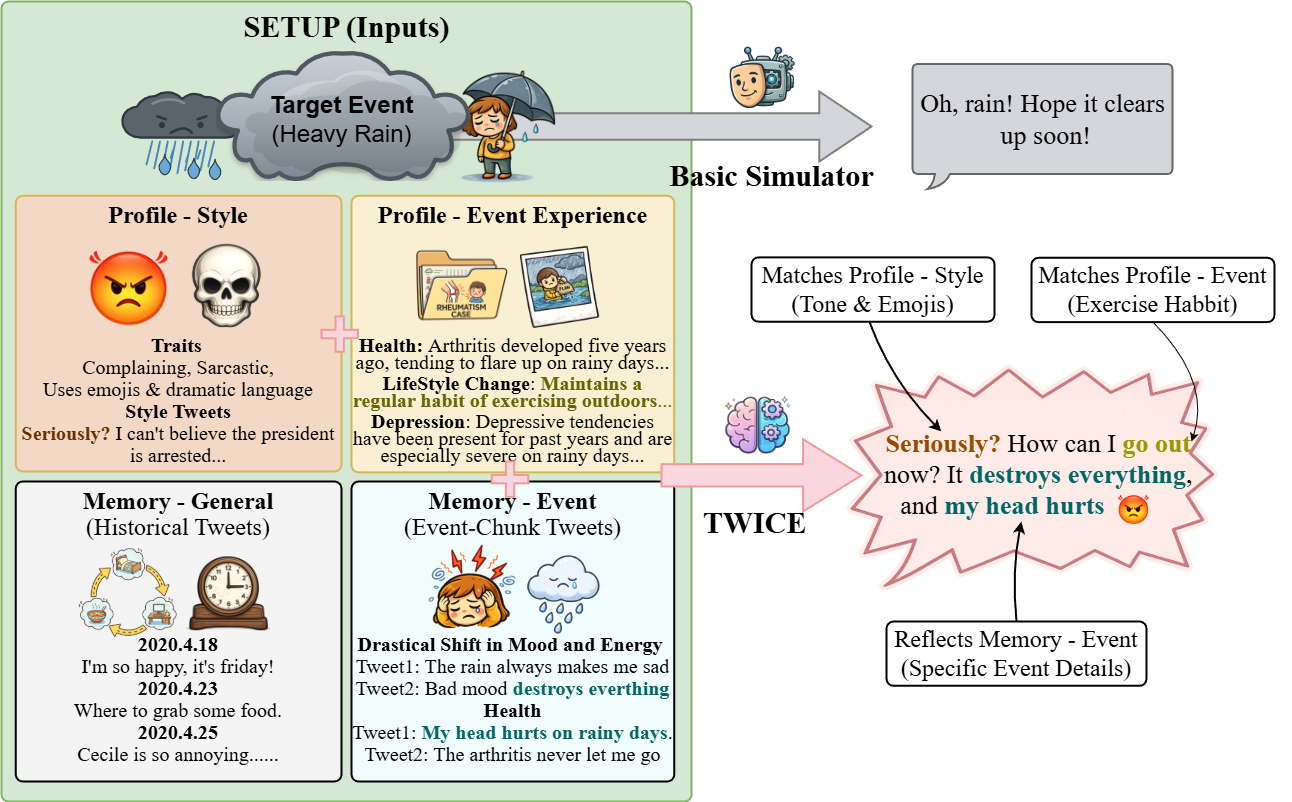}
    \caption{End-to-end example of \textsc{TWICE} on a real user-event instance, including extracted event summary, retrieved memories, first-stage generation, and style-rewritten output.}
    \label{fig:case_study}
    \vspace{-2mm}
\end{figure*}

By anchoring meaning in Stage I and enforcing idiolect in Stage II, the workflow produces simulations that are simultaneously personalized and time-aware. Figure~\ref{fig:case_study} shows an end-to-end example of \textsc{TWICE} for a real user-event instance.

\subsection{Multi-Faceted Evaluation Protocol}
To measure the fidelity of our simulated agents, we establish a comprehensive evaluation protocol that moves beyond simple string matching. We categorize our metrics into three pillars that directly address the core challenges of long-horizon personalization:

\textbf{Authenticity (Semantic Alignment)} evaluates how well the agent captures the ``what'' of the simulation by measuring sentence-level embedding semantic similarity (Sim) between the generated output and the ground-truth post. 

\textbf{Consistency (Stylistic Identity)} measures the ``how'' of the simulation. 
We measure style consistency by aggregating three complementary linguistic views:
(i) lexical similarity (TF-IDF), (ii) part-of-speech distribution, and (iii) sentence-length statistics.
This measures relative identity preservation rather than assuming that a user's style is perfectly static: given the same event context, a better simulator should retain more user-specific lexical, syntactic, and length-distributional patterns than non-personalized baselines.

Detailed definitions for each component are provided in Appendix~\ref{app:style-consistency}.


\begin{table*}[t]
\centering
\small
\setlength{\tabcolsep}{4.5pt}
\begin{tabular}{@{}l l c c c c c@{}}
\toprule
\textbf{Model} & \textbf{Baseline} &
\textbf{Sim $\uparrow$} &
\textbf{$|\Delta|$ FRE $\downarrow$} &
\textbf{$|\Delta|$ FKGL $\downarrow$} &
\textbf{Emotion KL $\downarrow$} &
\textbf{$\Delta$Tokens (Sim$-$Ori)} \\
\midrule

\multirow{7}{*}{ChatGPT-5.5}

& Memory-only & 0.7713 & 20.40 & 3.235 & \textbf{0.1942} & +17.76 \\

& Profile-only & 0.7373 & 21.11 & 3.832 & 0.2154 & +27.62 \\

& PopAlign persona & 0.7490 & 21.01 & 3.306 & 0.2718 & +26.90 \\

& OASIS & 0.7586 & 20.53 & 3.436 & 0.2130 & +20.43 \\

& Implicit profile & 0.7471 & 21.51 & 3.464 & 0.2286 & +18.88 \\

& TWICE (general) & 0.7789 & \textbf{20.12} & \textbf{3.220} & 0.2248 & +19.80 \\

& \textbf{TWICE (event)} & \textbf{0.7864} & 20.22 & 3.487 & 0.2319 & +21.93 \\

\midrule

\multirow{6}{*}{Qwen3-8B}
& Memory-only            & 0.7154 & 4.299  & \textbf{0.0292}  & 0.2663 & -4.53 \\
& PopAlign persona       & 0.6931 & 9.903  & 0.9826  & 0.4442 & +11.24 \\
& Implicit profile       & 0.7214 & \textbf{0.4020} & 0.6424  & \textbf{0.2443} & +3.34 \\
& OASIS                  & 0.7102 & 5.366  & 0.4148  & 0.3313 & +0.92 \\
& TWICE (general)        & 0.7215 & 5.746  & 1.110   & 0.3021 & +19.58 \\
& \textbf{TWICE (event)} & \textbf{0.7223} & 4.822  & 0.9456  & 0.2910 & +17.93 \\
\midrule

\multirow{8}{*}{Qwen3-14B}
& Non-personalized       & 0.7155 & 6.708  & 0.2941  & 0.2997 & -0.33 \\
& Memory-only            & 0.7201 & 3.210  & \textbf{0.2566}  & 0.2772 & -2.98 \\
& Profile-only           & 0.7097 & 9.204  & 0.9563  & 0.2918 & +6.94 \\
& PopAlign persona       & 0.6969 & 12.41  & 1.300   & 0.4137 & +11.51 \\
& Implicit profile       & 0.7247 & 3.198  & 0.4079  & \textbf{0.2562} & +8.90 \\
& OASIS                  & 0.7152 & 8.345  & 0.9226  & 0.2967 & +2.50 \\
& TWICE (general)        & 0.7205 & 1.412  & 0.8443  & 0.2914 & +13.57 \\
& \textbf{TWICE (event)} & \textbf{0.7275} & \textbf{1.102} & 0.7479 & 0.2959 & +12.93 \\
\bottomrule
\end{tabular}

\vspace{-2mm}
\caption{Baseline comparisons grouped by backbone models.
PopAlign persona~\citep{Hu2025PopAlign}: population-aligned persona conditioning.
Implicit profile~\citep{Wang2025USP}: USP-style implicit persona from long-term tweet history.
OASIS~\citep{Yang2025OASIS}: social-media agent baseline using coarse persona and recent posts.
$|\Delta|$ FRE/FKGL denote absolute readability deviations from ground truth (lower is better).
$\Delta$Tokens (Sim$-$Ori) denotes average token-count difference between simulated and original tweets.}
\label{tab:baseline}
\end{table*}

\textbf{Humanlikeness} evaluates whether the generated tweet exhibits natural readability patterns and affective structures aligned with the user's historical expression by measuring 1) \textbf{Readability}: absolute deviation ($|\Delta|$) in Flesch Reading Ease (FRE)~\cite{Flesch1948} and Flesch–Kincaid Grade Level (FKGL)~\cite{Kincaid1975} relative to the ground truth to ensure structural complexity matches the user's natural habits; 2) \textbf{Affective Complexity}: the KL divergence between Valence-Arousal-Dominance (VAD) distributions. A lower KL divergence indicates the simulated tweet replicates the user's emotional intensity and affective trajectory. Detailed definitions and formulas are provided in Appendix~\ref{app:humanlikeness}.

This protocol evaluates RQ1 (Personalization) through authenticity and style metrics, RQ2 (Temporal Coherence) through emotional alignment and sensitivity analyses, and RQ3 (Generalization) by comparing these results across high-dynamics and control cohorts.

\section{Experiments} \label{sec:experiments}

We evaluate how \textsc{TWICE} leverages personalized characteristics and long-term temporal patterns for authentic, consistent, and human-like tweet generation.

\subsection{Dataset and Data Processing}
To evaluate TWICE’s ability to model long-term temporal evolution, we utilize a large-scale longitudinal Twitter dataset containing 34,330 users~\cite{Suhavi2022TwitterSTMHDAE}. Detailed user-level statistics are provided in Appendix~\ref{app:dataset_stats}. To test the framework’s ability to model temporal evolution, we divided the dataset into two distinct cohorts:
\begin{itemize}
    \item \textbf{High-Dynamics Cohort (POS)}: A group of users (N = 26,131) associated with self-disclosed mental-health-related categories in the Twitter-STMHD dataset~\citep{Suhavi2022TwitterSTMHDAE}. We use this cohort as a high-dynamics setting because such self-disclosed categories are often accompanied by pronounced emotional and behavioral variation in longitudinal social-media activity. These users provide a rigorous testbed for our event-driven memory module's ability to track rapid temporal shifts.
    \item \textbf{Baseline Cohort (NEG)}: A control group of users (N = 8,199) without such reported markers, representing ``standard'' social media behavior with less pronounced fluctuations. This allows us to assess the framework’s generalizability to broader populations.
\end{itemize}

To extract features, we apply a multi-stage pipeline using the life-event taxonomy from PsyEvent~\cite{Chen2024MappingLC} and the behavioral schema from PsySym~\cite{Zhang2022SymptomIF}. These classifiers identify triggers (the ``why'') and emotional shifts (the ``how'') for each user. Only high-confidence tweets above a probability threshold are retained, after which an LLM performs structured event extraction (e.g., event type, emotion, and time expressions) for memory construction. Detailed descriptions of the classifier-based event detection and the subsequent structured extraction process are provided in Appendix~\ref{app:event_extraction}.

For diverse evaluation, we use Gaussian Kernel Density Estimation to sample 977 representative users across the semantic space. Finally, each user's timeline is split into priors (historical context for profiling/memory) and target events (the simulation objectives).

Our experimental analysis is aligned with the three research questions. RQ1 evaluates how TWICE compares with existing personalized user simulation baselines in generating authentic and temporally grounded tweets. RQ2 investigates how event-driven memory contributes to temporal coherence, evaluated through emotional alignment and retrieval sensitivity analyses. RQ3 tests whether TWICE generalizes across users with different levels of behavioral dynamics by comparing the High-Dynamics and Baseline cohorts. This unified setup allows us to validate both the technical components and the broader applicability of the framework.

\subsection{Methods of Comparison}
We compare TWICE against three categories of LLM-based simulators:
1) Generic Simulators, Non-personalized and Memory-only models that rely purely on recent context; 2) Profile-Centric Models, PopAlign persona~\cite{Hu2025PopAlign} and USP-style implicit profiling~\cite{Wang2025USP}  which use static persona grounding; 3) Social-Media Agents, The OASIS framework~\cite{Yang2025OASIS}, representing current state-of-the-art in multi-agent social simulation. Detailed sampling procedures, parameter settings, and baseline definitions are provided in Appendix \ref{app:experiment_settings}.
\begin{table}[htbp]
    \centering
    \small
    \setlength{\tabcolsep}{8pt}
    \renewcommand{\arraystretch}{1.3}
    \begin{tabular}{lccc}
        \toprule
        \textbf{Metric} & \textbf{Qwen3} & \textbf{DeepSeek} & \textbf{Llama} \\
        \midrule
        Similarity$\uparrow$ & 0.5270 & \cellcolor{gray!30}0.5274 & 0.5246 \\
        Style$\uparrow$ & \cellcolor{gray!30}0.3426 & 0.3395 & 0.3245 \\
        Word Overlap$\uparrow$ & 0.2205 & \cellcolor{gray!30}0.2342 & 0.2250 \\
        Emotion Intensity$\downarrow$ & \cellcolor{gray!30}0.0025 & 0.0279 & 0.0180 \\
        Readability Diff$\downarrow$ & \cellcolor{gray!30}0.4044 & 3.3186 & 0.5405 \\
        \bottomrule
    \end{tabular}
    \caption{Averaged Performance Comparison of Qwen3, DeepSeek, and Llama. Emotion Intensity and Readability are measured as differences relative to ground-truth.}
    \label{tab:avg_model_comparison}
\end{table}

\subsection{Implementation Details}
We implement TWICE across three backbone models\footnote{
Qwen3-8B~\cite{qwen3},
DeepSeek-R1-Distill-Qwen-7B~\cite{deepseekr1},
Llama-3.1-8B-Instruct~\cite{llama3herd}.}, with a primary focus on open-source LLMs due to considerations of cost, accessibility, and reproducibility. 
As shown in Table \ref{tab:avg_model_comparison}, Qwen3 outperformed others across key metrics (semantic similarity, style, emotion intensity). We therefore selected the Qwen model series for all subsequent experiments.
In addition, we include ChatGPT-5.5~\cite{openai2026gpt55} as a closed-source reference model in Table~\ref{tab:baseline} to provide an external performance benchmark. This comparison helps contextualize the performance of TWICE relative to high-performing proprietary systems, while our main analysis remains centered on open-source backbones.
The event-driven memory module uses a one-year retrieval window and a memory size of $N=10$, which provide the best trade-off between contextual grounding and noise according to the sensitivity analysis in Section~\ref{main:4.2}. 
All other factors including prompts, decoding settings, the \textsc{TWICE} pipeline, temporal/memory parameters, and evaluation data were held constant to isolate LLM performance.



\section{Results and Analysis}\label{sec:results}

We present our results structured by our three research questions, focusing on the framework’s overall effectiveness, the impact of its temporal components, and its performance across user cohorts.

\subsection{Main Performance Results (RQ1)}
To address \textbf{RQ1}, we compare TWICE against state-of-the-art baselines. As shown in Table~\ref{tab:baseline}, TWICE achieves consistently strong performance across semantic authenticity, readability alignment, and temporal personalization metrics across backbones.

\paragraph{Authenticity and Style.} TWICE achieves highest semantic similarity and stylistic alignment. The two-stage workflow grounds content in Stage I and applies personalized style adaptation in Stage II.

\paragraph{Humanlikeness.} TWICE achieves strong readability alignment while maintaining competitive Emotion KL divergence, indicating stronger stylistic naturalness and affective consistency.

Comparisons highlight the necessity of modeling both a stable persona and event-conditioned temporal memory. While the Implicit profile (USP-style) baseline improves over simpler variants, it underperforms \textsc{TWICE}, showing that an implicit persona alone can't capture fine-grained behavioral dynamics. Similarly, PopAlign persona and OASIS exhibit limited gains, confirming that generic or recent-history-driven approaches are insufficient for simulating long-term, event-related behavior.

\paragraph{Human Evaluation.}
To complement the automatic metrics, we conduct a human evaluation on the NEG-cohort generalization setting using 50 sampled user--event instances. Annotators rank anonymized generations from TWICE and two representative personalization baselines along three dimensions: consistency, authenticity, and humanlikeness. Three annotators participated in the study, yielding 450 ranked judgments with moderate inter-annotator agreement (mean pairwise Kendall's $\tau = 0.401$). 

Results are summarized in Table~\ref{tab:human_eval}. TWICE achieves the best mean rank across all evaluation dimensions, indicating stronger behavioral consistency and perceived realism than both baselines. Additional evaluation details, agreement statistics, and pairwise analyses are provided in Appendix~\ref{appendix:human_eval}.

\begin{table}[t]
\centering
\small
\setlength{\tabcolsep}{3pt}
\begin{tabular}{lccc}
\toprule
Method & Consist. & Auth. & Human. \\
\midrule
Non-personalized & 2.28 & 2.14 & 2.05 \\
Implicit Profile & 1.93 & 1.99 & 2.05 \\
\textbf{TWICE (ours)} & \textbf{1.79} & \textbf{1.87} & \textbf{1.90} \\
\bottomrule
\end{tabular}
\caption{Per-dimension mean ranks in human evaluation (lower is better). 
Consist., Auth., and Human. denote consistency, authenticity, and humanlikeness.}
\label{tab:human_eval}
\end{table}

\subsection{Validation and Sensitivity of Event-Driven Memory (RQ2)}
\label{main:4.2}

RQ2 examines the contribution of the event-driven memory module and how different retrieval design choices affect performance. Our results show that event-centered memory consistently improves temporally grounded personalization, particularly when combined with structured event-aware user profiles.

\paragraph{Event-centered memory provides stronger temporal grounding.}

The ablation results (Table~\ref{tab:ablation_all}) show that event-driven memory consistently outperforms general memory. This indicates that organizing past tweets around detected life events provides more informative contextual grounding than relying on coarse temporal aggregation alone. Furthermore, event-aware user profiles achieve stronger results than general profiles or no-profile settings, suggesting that stable personal priors and event-grounded memory play complementary roles in modeling user behavior.

\begin{table}[t]
\centering
\footnotesize
\setlength{\tabcolsep}{5pt}
\begin{tabular}{lcc}
\toprule
\textbf{Setting} 
& \textbf{Sim $\uparrow$} 
& \textbf{$|\Delta|$ FKGL $\downarrow$} \\
\midrule
\multicolumn{3}{l}{\textit{Memory ablation (no profile):}} \\
General memory            & 0.7160 & 3.8805 \\
Event memory              & \textbf{0.7306} & \textbf{3.5921} \\
\midrule
\multicolumn{3}{l}{\textit{Profile ablation (general memory):}} \\
No profile                & 0.7160 & 3.8805 \\
General profile           & 0.7135 & 3.7513 \\
Event profile             & \textbf{0.7267} & \textbf{3.6859} \\
\midrule
\multicolumn{3}{l}{\textit{Retrieval scoring ablation:}} \\
Sim-only                  & 0.7205 & 0.9612 \\
Recency-only              & 0.7134 & 3.7623 \\
Sim$\times$Time           & 0.7249 & 3.8063 \\
Full scoring (\textsc{TWICE}) & \textbf{0.7305} & \textbf{0.5653} \\
\bottomrule
\end{tabular}
\vspace{-2mm}
\caption{Ablation study of memory/profile construction and retrieval scoring variants.  $|\Delta|$ FKGL is the absolute readability deviation from ground‑truth posts.}
\label{tab:ablation_all}
\end{table}

\paragraph{Moderate memory retrieval improves coherence.}

\begin{figure}[!htbp]
    \centering
    \includegraphics[width=1\linewidth]{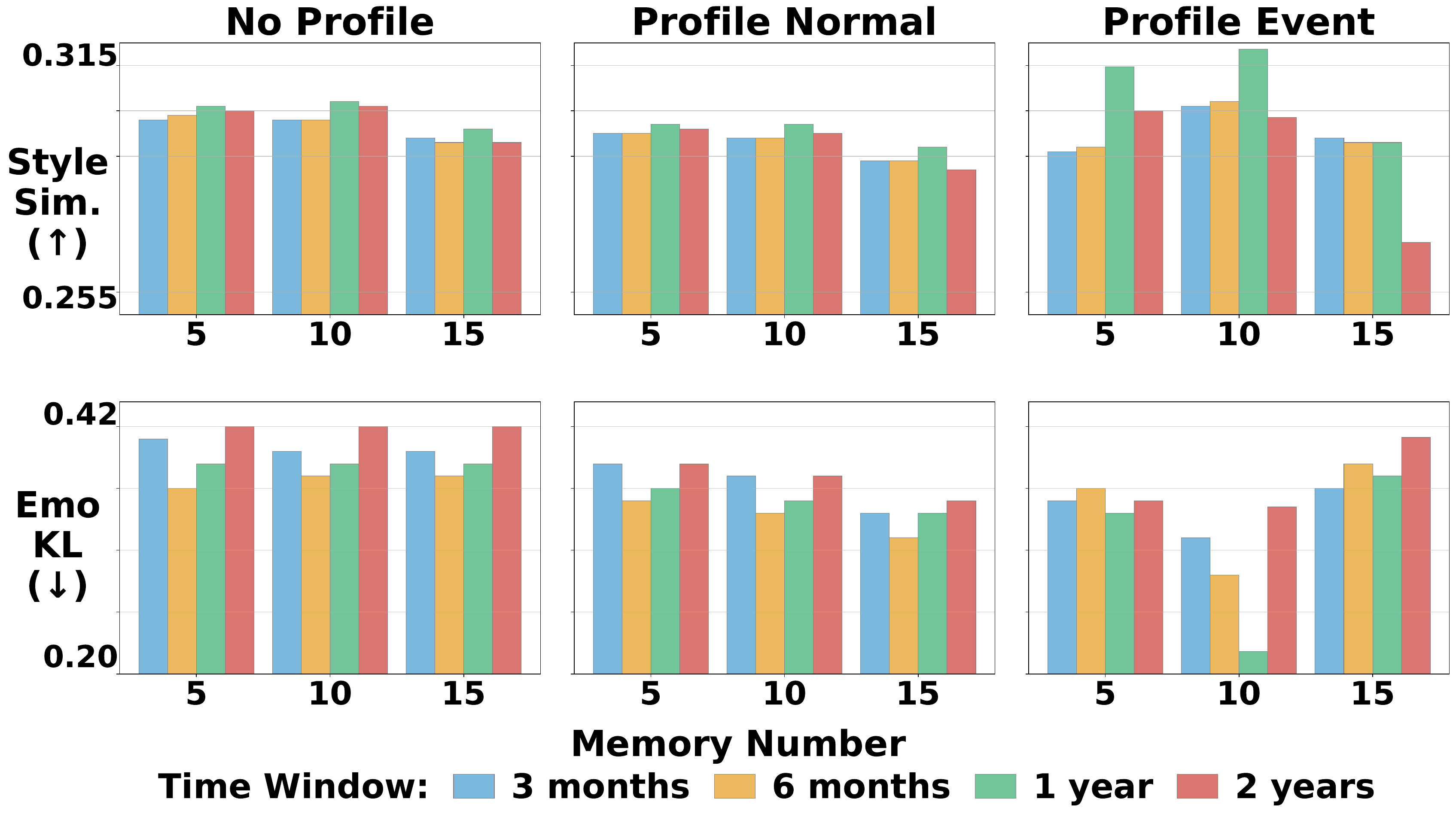}
    \caption{Sensitivity analysis of TWICE with respect to profile construction, retrieval window, and memory size. We compare different configurations across style similarity and emotion alignment. TWICE achieves the best performance under moderate memory sizes and intermediate retrieval windows, indicating that excessive retrieval introduces noisy or irrelevant context.}
    \label{fig:temporal}
\end{figure}

We further analyze how retrieval design influences performance (Figure~\ref{fig:temporal}). Moderate memory numbers (5--10) consistently outperform larger memory pools, suggesting that excessive retrieval may introduce noise rather than improve coherence. Similarly, medium retrieval windows—particularly around one year—provide the best balance between long-term contextual grounding and outdated historical signals.

\paragraph{Effective retrieval requires both relevance and time.}

Finally, we examine different memory scoring strategies (Table~\ref{tab:ablation_all}). Similarity-only retrieval achieves strong semantic alignment but often ignores temporal locality, whereas recency-only retrieval improves temporal relevance at the cost of semantic consistency. Combining similarity with temporal decay yields the most stable results, confirming that effective temporal modeling requires jointly considering both what happened and when it happened.

\subsection{Behavioral Dynamics and Generalization (RQ3)}
Finally, we address RQ3 by evaluating whether TWICE generalizes across users with different levels of behavioral dynamics, as shown in Table \ref{tab:comp}.
For the High-Dynamics Cohort (POS), where users exhibit more pronounced behavioral fluctuations associated with self-disclosed mental-health-related signals, TWICE shows its largest margin of improvement. The event-driven module is particularly effective at tracking shifting emotional trajectories during periods of stronger affective variation.

\begin{figure}[!htbp]
    \centering
    \includegraphics[width=1\linewidth]{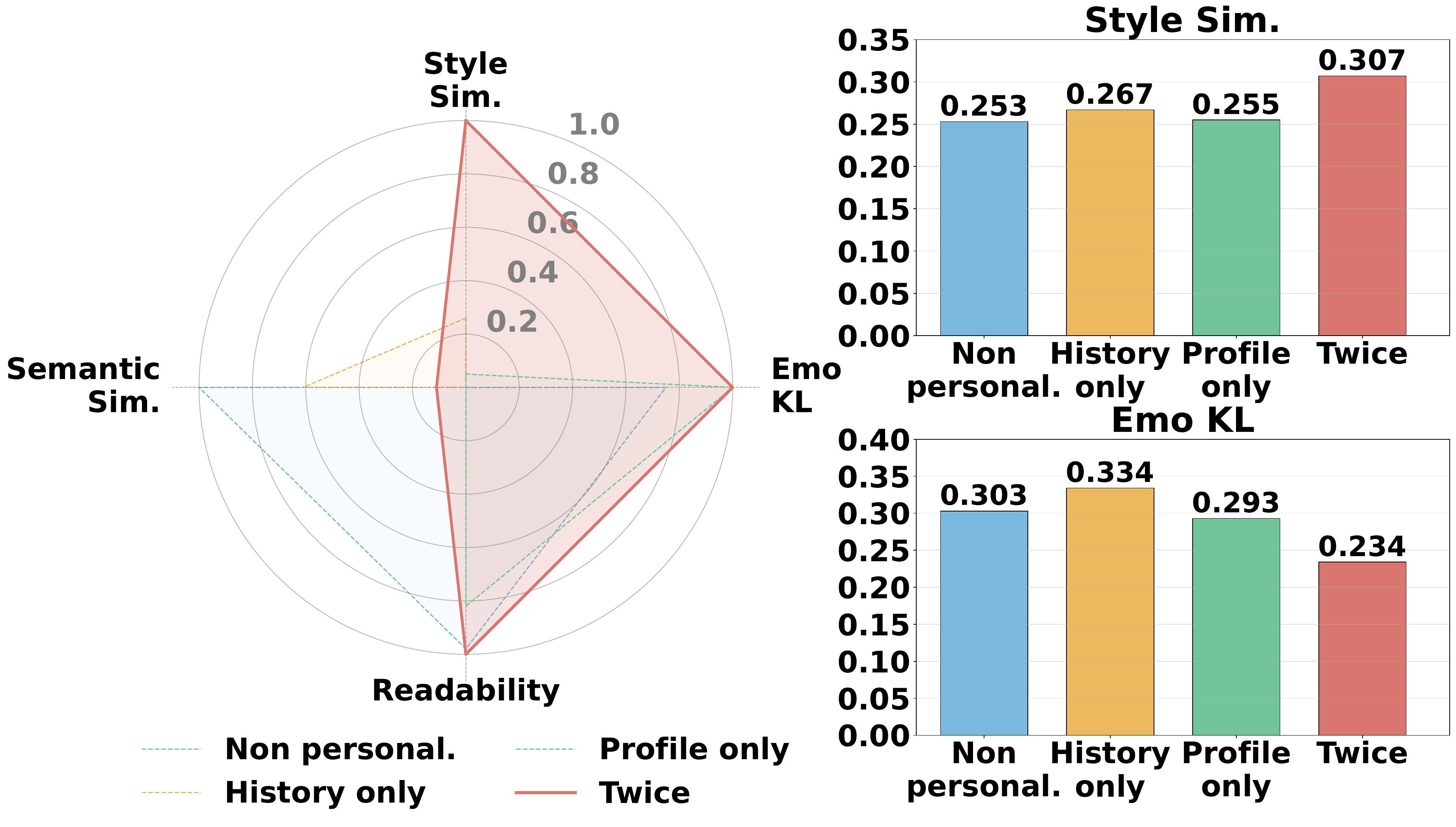}
    \caption{Generalization analysis on control users (NEG). We compare different simulation strategies (Non-personalized, History-only, Profile-only, and TWICE) across style similarity and emotion alignment. TWICE (event) consistently achieves the best trade-off between the two metrics, demonstrating strong generalization to users with stable behavioral patterns.}
    \label{fig:neg_generalization}
\end{figure}

The robustness of TWICE extends to the baseline cohort (NEG), as illustrated in Figure \ref{fig:neg_generalization}. Even for users with stable, low-variance behavior, TWICE maintains superior stylistic consistency and the lowest emotion KL divergence among the four representative strategies. This confirms that the framework’s gains are not contingent on cohort-specific affective signals or extreme emotional trajectories. Instead, TWICE captures fundamental patterns of temporally grounded behavior that generalize effectively to ordinary users with subtle behavioral fluctuations.

\begin{table}[htbp]
    \centering
    \small
    \setlength{\tabcolsep}{3pt}
    \renewcommand{\arraystretch}{1.3}
    \begin{tabular}{lccccc}
        \toprule
        \textbf{Cat.} & \textbf{Emo.$\downarrow$} & \textbf{Style$\uparrow$} & \textbf{$|\Delta|$ FRE $\downarrow$} & \textbf{$|\Delta|$ FKGL $\downarrow$} & \textbf{Sim.$\uparrow$} \\
        \midrule
        NEG & 0.3184 & 0.3192 & 7.4792 & 1.5551 & 0.4434 \\
        POS & \cellcolor{gray!30}0.2640 & \cellcolor{gray!30}0.3732 & \cellcolor{gray!30}0.0997 & \cellcolor{gray!30}0.8822 & \cellcolor{gray!30}0.4724 \\
        \bottomrule
    \end{tabular}
    \caption{
    Performance comparison across the NEG and POS cohorts.
    Metrics include Emotion (KL divergence; denoted as Emo.),
    Style similarity,
    Readability (absolute differences in FRE and FKGL relative to the original tweets),
    and Semantic similarity (denoted as Sim.).
    }
    \label{tab:comp}
\end{table}


\subsection{Qualitative Case Study}
As shown in Figure~\ref{fig:case_study}, a qualitative look at the generation stages reveals how TWICE handles complex context. In this instance, Stage I identifies the core event and emotional valence (e.g., frustration with a specific life event), while Stage II injects the user's specific linguistic habits, such as preferred slang and idiosyncratic punctuation, resulting in a simulation that is indistinguishable from the ground truth.

\section{Related Work}
\paragraph{User simulation.}

User simulation has been widely studied in interactive AI systems, ranging from early rule-based dialogue simulators \citep{Schatzmann2006Survey, Schatzmann2007AgendaBasedUS} to neural sequence models \citep{Sutskever2014Sequence}. Recent LLM-based approaches enable more realistic user behavior generation conditioned on historical interactions and contextual signals \citep{Brown2020FewShot, Kong2023LargeLM, Wang2023UserBS}, and have been extended to recommendation and social-network environments \citep{Wang2023RecAgent, Gao2023S3, Yang2025OASIS}. However, these approaches mainly focus on short interaction sessions or macro-level social dynamics, rather than modeling how individual behavior evolves over long temporal horizons.

\paragraph{Persona and personalization modeling.}

Persona-based or profile-based approaches improve personalization by inferring relatively stable user traits from historical interactions and generating responses aligned with user identity and preferences \citep{Wang2025USP, Hu2025PopAlign}. These approaches improve user alignment and stylistic consistency, but typically treat user characteristics as static or slowly changing priors, without modeling how behavioral responses evolve across events over time.

\paragraph{Memory mechanisms in LLM agents.}
Recent work has shown that explicit memory is critical for extending the reasoning and behavioral coherence of LLM-based agents. Generative Agents \citep{Park2023GenerativeAgents} demonstrate that storing and retrieving past observations can support coherent agent behavior in simulated environments, while multi-agent frameworks such as CAMEL \citep{Li2023CAMEL} further highlight the importance of contextual memory in sustaining role-consistent interactions. More generally, memory-augmented LLM agents use stored histories to improve consistency, planning, and long-context reasoning. However, these memory mechanisms primarily organize conversational context, observations, or interaction traces, rather than temporally structured life events and their associated behavioral trajectories.

\paragraph{Temporal and event-grounded behavior modeling.}

Prior work in social media and mental-health NLP has shown that online expressions reflect long-term psychological states and affective changes \citep{Chancellor2020MethodsIP, Naslund2020SocialMA, DeChoudhury2013PredictingDepression}. More recent studies further highlight the role of life events and symptom-related signals in shaping users' temporal behavior online \citep{Chen2024MappingLC, Zhang2022SymptomIF}. These findings suggest that realistic user simulation should account not only for who the user is, but also for what events they have experienced and how those experiences shape later expression. Nevertheless, existing LLM-based user simulators rarely incorporate event-centered memory structures that explicitly connect past experiences to present reactions. Compared with agent-based simulations such as Generative Agents~\citep{Park2023GenerativeAgents}, TWICE explicitly organizes memory around life events and models temporally grounded behavioral rather than generic contextual recall.

\section{Conclusion}

We presented TWICE, a temporally grounded framework for personalized user simulation that models not only what users say, but also how past events shape later expression over time. By combining structured user profiles, event-driven memory, and staged style adaptation, TWICE captures both stable identity traits and evolving behavioral dynamics in social media contexts. Experiments show that TWICE consistently improves the authenticity, consistency, emotional alignment, and humanlikeness of simulated tweets, particularly under stronger temporal and affective variation.

More broadly, our findings suggest that temporally grounded user simulation can serve as a useful behavioral modeling framework for studying long-term human expression and interaction dynamics. We hope this work motivates future research on personalized agents, social simulation, and human-centered generative AI with richer temporal and behavioral grounding.

\section*{Limitations}
Our work has several limitations. First, our evaluation primarily relies on automatic metrics such as semantic similarity, stylistic similarity, readability differences, and emotion divergence. While these metrics enable scalable comparisons across large numbers of simulated tweets, they cannot fully capture nuanced human judgments of authenticity, contextual appropriateness, or perceived humanlikeness. Future work could complement these measures with controlled human evaluations or expert assessments.

Second, TWICE depends on an upstream event detection pipeline that combines supervised classifiers with LLM-based structured extraction. Although confidence filtering reduces spurious signals, errors in event identification or ambiguous expressions in social media text may propagate to the memory construction stage and affect downstream simulation quality.

Finally, TWICE models temporally grounded behavioral patterns through retrieved memories but does not explicitly represent the causal mechanisms underlying behavioral changes. The framework therefore captures correlations between past events and later expressions rather than the psychological or social processes driving them. Incorporating causal modeling or richer behavioral representations remains an important direction for future work.

\section*{Ethical Considerations}

Our study involves user-level simulation on social media, including users associated with self-disclosed mental-health-related categories. Although we only use publicly available tweets from previously released datasets, generating temporally adaptive personalized content may still introduce privacy and misuse risks, such as unintended re-identification, sensitive attribute leakage, or malicious impersonation. We therefore report only aggregated statistics, avoid releasing generated user-level outputs, and restrict qualitative examples to anonymized and de-identified cases.

Our framework is designed for research purposes (e.g., controlled evaluation, benchmarking, and stress-testing) rather than real-world deployment. In particular, TWICE should not be used for clinical decision-making or for generating content intended to influence individuals, as simulated outputs may omit important context or contain plausible but inaccurate cues. Any future deployment would require substantially stronger privacy and safety safeguards.

More fundamentally, we emphasize that temporally adaptive user simulation is intended as a controlled research capability for studying language use and behavioral dynamics, rather than impersonating real individuals in online settings. Similar simulation methodologies are widely used in computational social science, human-centered NLP, and agent-based modeling to study interaction patterns, temporal variation, and behavioral adaptation under controlled conditions. TWICE is not intended for deployment in automated persuasion, clinical intervention, or identity impersonation systems. These considerations motivate clear usage boundaries, de-identification, and aggregate-level reporting practices throughout this work.

\bibliography{aaai2026}

@article{Chancellor2020MethodsIP,
  title={Methods in predictive techniques for mental health status on social media: a critical review},
  author={Stevie Chancellor and Munmun De Choudhury},
  journal={NPJ Digital Medicine},
  year={2020},
  volume={3}
}

@article{Wang2023UserBS,
author = {Wang, Lei and Zhang, Jingsen and Yang, Hao and Chen, Zhi-Yuan and Tang, Jiakai and Zhang, Zeyu and Chen, Xu and Lin, Yankai and Sun, Hao and Song, Ruihua and Zhao, Xin and Xu, Jun and Dou, Zhicheng and Wang, Jun and Wen, Ji-Rong},
title = {User Behavior Simulation with Large Language Model-based Agents},
year = {2025},
issue_date = {March 2025},
publisher = {Association for Computing Machinery},
address = {New York, NY, USA},
volume = {43},
number = {2},
issn = {1046-8188},
url = {https://doi.org/10.1145/3708985},
doi = {10.1145/3708985},
journal = {ACM Trans. Inf. Syst.},
month = jan,
articleno = {55},
numpages = {37}
}

@inproceedings{Wang2025USP,
  author = {Wang, Kuang and Li, Xianfei and Yang, Shenghao and Zhou, Li and Jiang, Feng and Li, Haizhou},
  title = {Know You First and Be You Better: Modeling Human-Like User Simulators via Implicit Profiles},
  booktitle = {Proceedings of the 63rd Annual Meeting of the Association for Computational Linguistics (ACL)},
  year = {2025}
}

@inproceedings{Chen2024MappingLC,
    title = "Mapping Long-term Causalities in Psychiatric Symptomatology and Life Events from Social Media",
    author = "Chen, Siyuan  and
      Wang, Meilin  and
      Lv, Minghao  and
      Zhang, Zhiling  and
      Ju, Qianqian  and
      Dejiyangla  and
      Peng, Yujia  and
      Zhu, Kenny Q.  and
      Wu, Mengyue",
    editor = "Duh, Kevin  and
      Gomez, Helena  and
      Bethard, Steven",
    booktitle = "Proceedings of the 2024 Conference of the North American Chapter of the Association for Computational Linguistics: Human Language Technologies (Volume 1: Long Papers)",
    month = jun,
    year = "2024",
    address = "Mexico City, Mexico",
    publisher = "Association for Computational Linguistics",
    url = "https://aclanthology.org/2024.naacl-long.306/",
    doi = "10.18653/v1/2024.naacl-long.306",
    pages = "5472--5487"
}

@article{Kong2023LargeLM,
  title={Large Language Model as a User Simulator},
  author={Chuyi Kong and Yaxin Fan and Xiang Wan and Feng Jiang and Benyou Wang},
  journal={ArXiv},
  year={2023},
  volume={abs/2308.11534}
}

@article{Naslund2020SocialMA,
  title={Social Media and Mental Health: Benefits, Risks, and Opportunities for Research and Practice},
  author={John A. Naslund and Ameya P. Bondre and John B Torous and Kelly A. Aschbrenner},
  journal={Journal of Technology in Behavioral Science},
  year={2020},
  volume={5},
  pages={245 - 257}
}

@inproceedings{Schatzmann2007AgendaBasedUS,
  title={Agenda-Based User Simulation for Bootstrapping a POMDP Dialogue System},
  author={Jost Schatzmann and Blaise Thomson and Karl Weilhammer and Hui Ye and Steve J. Young},
  booktitle={North American Chapter of the Association for Computational Linguistics},
  year={2007}
}

@inproceedings{Brown2020FewShot,
  author       = {Tom B. Brown and
                  Benjamin Mann and
                  Nick Ryder and
                  Melanie Subbiah and
                  Jared Kaplan and
                  Prafulla Dhariwal and
                  Arvind Neelakantan and
                  Pranav Shyam and
                  Girish Sastry and
                  Amanda Askell and
                  Sandhini Agarwal and
                  Ariel Herbert{-}Voss and
                  Gretchen Krueger and
                  Tom Henighan and
                  Rewon Child and
                  Aditya Ramesh and
                  Daniel M. Ziegler and
                  Jeffrey Wu and
                  Clemens Winter and
                  Christopher Hesse and
                  Mark Chen and
                  Eric Sigler and
                  Mateusz Litwin and
                  Scott Gray and
                  Benjamin Chess and
                  Jack Clark and
                  Christopher Berner and
                  Sam McCandlish and
                  Alec Radford and
                  Ilya Sutskever and
                  Dario Amodei},
  editor       = {Hugo Larochelle and
                  Marc'Aurelio Ranzato and
                  Raia Hadsell and
                  Maria{-}Florina Balcan and
                  Hsuan{-}Tien Lin},
  title        = {Language Models are Few-Shot Learners},
  booktitle    = {Advances in Neural Information Processing Systems 33: Annual Conference
                  on Neural Information Processing Systems 2020, NeurIPS 2020, December
                  6-12, 2020, virtual},
  year         = {2020},
  url          = {https://proceedings.neurips.cc/paper/2020/hash/1457c0d6bfcb4967418bfb8ac142f64a-Abstract.html},
  bibsource    = {dblp computer science bibliography, https://dblp.org}
}

@inproceedings{Zhang2022SymptomIF,
  title={Symptom Identification for Interpretable Detection of Multiple Mental Disorders on Social Media},
  author={Zhiling Zhang and Siyuan Chen and Mengyue Wu and Ke Zhu},
  booktitle={Conference on Empirical Methods in Natural Language Processing},
  year={2022}
}

@inproceedings{Suhavi2022TwitterSTMHDAE,
  title={Twitter-STMHD: An Extensive User-Level Database of Multiple Mental Health Disorders},
  author={Suhavi and Asmit Kumar Singh and Udit Arora and Somyadeep Shrivastava and Aryaveer Singh and Rajiv Ratn Shah and Ponnurangam Kumaraguru},
  booktitle={International Conference on Web and Social Media},
  year={2022}
}

@article{article,
author = {Tanana, Michael and Soma, Christina and Srikumar, Vivek and Atkins, David and Imel, Zac},
year = {2019},
month = {04},
pages = {},
title = {Development and evaluation of ClientBot: A patient- like conversational agent to train basic counseling skills},
volume = {21},
journal = {Journal of Medical Internet Research},
doi = {10.2196/12529}
}

@inproceedings{Park2023GenerativeAgents,
author = {Park, Joon Sung and O'Brien, Joseph and Cai, Carrie Jun and Morris, Meredith Ringel and Liang, Percy and Bernstein, Michael S.},
title = {Generative Agents: Interactive Simulacra of Human Behavior},
year = {2023},
isbn = {9798400701320},
publisher = {Association for Computing Machinery},
address = {New York, NY, USA},
url = {https://doi.org/10.1145/3586183.3606763},
doi = {10.1145/3586183.3606763},
booktitle = {Proceedings of the 36th Annual ACM Symposium on User Interface Software and Technology},
articleno = {2},
numpages = {22},
location = {San Francisco, CA, USA},
series = {UIST '23}
}

@inproceedings{Li2023CAMEL,
  author = {Li, Guohao and Hammoud, Hasan Abed Al Kader and Itani, Hani and Khizbullin, Dmitrii and Ghanem, Bernard},
  title = {CAMEL: Communicative Agents for Mind Exploration of Large Language Model Society},
  booktitle = {Advances in Neural Information Processing Systems (NeurIPS)},
  year = {2023}
}

@inproceedings{DeChoudhury2013PredictingDepression,
author = {De Choudhury, Munmun and Gamon, Michael and Counts, Scott and Horvitz, Eric},
title = {Predicting Depression via Social Media},
booktitle = {Proceedings of the 7th International AAAI Conference on Web and Social Media (ICWSM)},
year = {2013},
pages = {128--137},
publisher = {AAAI}
}

@inproceedings{Sutskever2014Sequence,
  author = {Sutskever, Ilya and Vinyals, Oriol and Le, Quoc V.},
  title = {Sequence to Sequence Learning with Neural Networks},
  booktitle = {Advances in Neural Information Processing Systems (NeurIPS)},
  pages = {3104--3112},
  year = {2014},
  publisher = {Curran Associates, Inc.}
}

@inproceedings{Wang2023RecAgent,
  author = {Wang, Yifan and Zhang, Jiarui and Jiang, Yuxuan and Li, Yuyuan and Sun, Aixin},
  title = {RecAgent: A Novel Simulation Paradigm for Recommender Systems},
  booktitle = {Proceedings of the ACM International Conference on Web Search and Data Mining (WSDM)},
  year = {2024}
}

@article{Gao2023S3,
  author = {Gao, Chen and Lan, Xiaochong and Lu, Zhijie and Mao, Jinzhu and Piao, Jing and Wang, Huandong and Jin, Depeng and Li, Yong},
  title = {S3: Social-Network Simulation System with Large Language Model Empowered Agents},
  journal = {arXiv preprint arXiv:2307.14984},
  year = {2023}
}

@article{Flesch1948,
  title={A new readability yardstick},
  author={Flesch, Rudolf},
  journal={Journal of Applied Psychology},
  volume={32},
  number={3},
  pages={221--233},
  year={1948}
}

@techreport{Kincaid1975,
  title={Derivation of new readability formulas for Navy enlisted personnel},
  author={Kincaid, J. Peter and Fishburne, Robert and Rogers, Richard and Chissom, Brad},
  institution={Naval Technical Training Command},
  year={1975}
}

@misc{li2024evolvingagentsinteractivesimulation,
      title={Evolving Agents: Interactive Simulation of Dynamic and Diverse Human Personalities}, 
      author={Jiale Li and Jiayang Li and Jiahao Chen and Yifan Li and Shijie Wang and Hugo Zhou and Minjun Ye and Yunsheng Su},
      year={2024},
      eprint={2404.02718},
      archivePrefix={arXiv},
      primaryClass={cs.HC},
      url={https://arxiv.org/abs/2404.02718}, 
}

@book{Ricci2015RecommenderHandbook,
  title     = {Recommender Systems Handbook},
  editor    = {Ricci, Francesco and Rokach, Lior and Shapira, Bracha},
  edition   = {2nd},
  publisher = {Springer US},
  address   = {Boston, MA},
  year      = {2015},
  isbn      = {978-1-4899-7636-9},
  doi       = {10.1007/978-1-4899-7637-6},
  url       = {https://link.springer.com/book/10.1007/978-1-4899-7637-6}
}

@article{Yang2025OASIS,
  author       = {Ziyi Yang and
                  Zaibin Zhang and
                  Zirui Zheng and
                  Yuxian Jiang and
                  Ziyue Gan and
                  Zhiyu Wang and
                  Zijian Ling and
                  Jinsong Chen and
                  Martz Ma and
                  Bowen Dong and
                  Prateek Gupta and
                  Shuyue Hu and
                  Zhenfei Yin and
                  Guohao Li and
                  Xu Jia and
                  Lijun Wang and
                  Bernard Ghanem and
                  Huchuan Lu and
                  Chaochao Lu and
                  Wanli Ouyang and
                  Yu Qiao and
                  Philip Torr and
                  Jing Shao},
  title        = {{OASIS:} Open Agent Social Interaction Simulations with One Million
                  Agents},
  journal      = {CoRR},
  volume       = {abs/2411.11581},
  year         = {2024},
  url          = {https://doi.org/10.48550/arXiv.2411.11581},
  doi          = {10.48550/ARXIV.2411.11581},
  eprinttype    = {arXiv},
  eprint       = {2411.11581},
  bibsource    = {dblp computer science bibliography, https://dblp.org}
}

@misc{Hu2025PopAlign,
      title={Population-Aligned Persona Generation for LLM-based Social Simulation}, 
      author={Zhengyu Hu and Jianxun Lian and Zheyuan Xiao and Max Xiong and Yuxuan Lei and Tianfu Wang and Kaize Ding and Ziang Xiao and Nicholas Jing Yuan and Xing Xie},
      year={2025},
      eprint={2509.10127},
      archivePrefix={arXiv},
      primaryClass={cs.CL},
      url={https://arxiv.org/abs/2509.10127}, 
}

@misc{qwen3,
  title        = {Qwen3 Technical Report},
  author       = {Qwen Team},
  year         = {2025},
  eprint       = {2505.09388},
  archivePrefix= {arXiv},
  primaryClass = {cs.CL}
}

@misc{deepseekr1,
  title        = {DeepSeek-R1: Incentivizing Reasoning Capability in LLMs via Reinforcement Learning},
  author       = {DeepSeek-AI},
  year         = {2025},
  eprint       = {2501.12948},
  archivePrefix= {arXiv},
  primaryClass = {cs.CL}
}

@misc{llama3herd,
  title        = {The Llama 3 Herd of Models},
  author       = {Meta AI},
  year         = {2024},
  eprint       = {2407.21783},
  archivePrefix= {arXiv},
  primaryClass = {cs.CL}
}

@article{Schatzmann2006Survey,
author = {Schatzmann, Jost and Weilhammer, Karl and Stuttle, Matthew and Young, Steve},
year = {2006},
month = {06},
pages = {97-126},
title = {A survey of statistical user simulation techniques for reinforcement-learning of dialogue management strategies},
volume = {21},
journal = {Knowledge Eng. Review},
doi = {10.1017/S0269888906000944}
}

@article{Lazer2009ComputationalSocialScience,
  title        = {Computational Social Science},
  author       = {Lazer, David and Pentland, Alex and Adamic, Lada and Aral, Sinan and Barab{\'a}si, Albert-L{\'a}szl{\'o} and Brewer, Devon and Christakis, Nicholas and Contractor, Noshir and Fowler, James and Gutmann, Myron and Jebara, Tony and King, Gary and Macy, Michael and Roy, Deb and Van Alstyne, Marshall},
  journal      = {Science},
  volume       = {323},
  number       = {5915},
  pages        = {721--723},
  year         = {2009},
  publisher    = {American Association for the Advancement of Science},
  doi          = {10.1126/science.1167742},
  url          = {https://www.science.org/doi/10.1126/science.1167742}
}

@misc{openai2026gpt55,
  author       = {{OpenAI}},
  title        = {GPT-5.5 System Card},
  year         = {2026},
  howpublished = {\url{https://openai.com/index/gpt-5-5-system-card/}},
  note         = {Accessed: 2026-05-22}
}

\clearpage
\appendix

\section{Additional Dataset Statistics}
\label{app:dataset_stats}

\begin{table}[htbp]
\centering
\small
\setlength{\tabcolsep}{2pt}
\renewcommand{\arraystretch}{1.2}
\begin{tabular}{cccc}
\toprule
\textbf{Category} & \textbf{\# Users} & \textbf{Avg. posts} & \textbf{Avg. time span (days)} \\
\midrule
ADHD & 8{,}095 & 7{,}592.32 & 1{,}325.52 \\
Anxiety & 4{,}843 & 10{,}190.79 & 1{,}801.22 \\
Bipolar & 1{,}651 & 11{,}639.16 & 2{,}246.95 \\
Depression & 6{,}803 & 7{,}766.64 & 1{,}141.31 \\
NEG & 8{,}199 & 9{,}845.13 & 1{,}424.66 \\
OCD & 1{,}325 & 6{,}597.84 & 995.26 \\
PTSD & 3{,}414 & 7{,}410.33 & 1{,}096.22 \\
\midrule
\textbf{All (weighted)} & \textbf{34{,}330} & \textbf{8{,}669.61} & \textbf{1{,}388.57} \\
\bottomrule
\end{tabular}
\caption{User-level statistics by category. The “All (weighted)” row reports user-count–weighted averages across categories for average posts per user and average time span (days).}
\label{tab:category_stats}
\end{table}

\section{Feature Extraction for User Profiling}
\label{app:profile}

To construct structured user profiles, we extract both general attributes and personalized behavioral signals from historical tweets.

\subsection{General attributes} 
Demographic attributes such as age, gender, and career are extracted using a hybrid pipeline that combines regular-expression patterns with embedding-based semantic matching to handle linguistic variability. Lightweight LLM inference is applied when contextual disambiguation is required.

\subsection{Personalized behavioral attributes} 
To capture identity-level regularities and temporal behavioral signals, we incorporate life events, symptom cues, personality traits, and stylistic evidence. Following Chen et al.~\cite{Chen2024MappingLC} and Zhang et al.~\cite{Zhang2022SymptomIF}, we use their supervised classifiers to obtain life-event and symptom relevance scores for each tweet. High-confidence signals are grouped by event type and summarized into compact descriptions using an LLM.

\subsection{Personality and style} 
Big Five traits are inferred from historical posts using LLM-based analysis. For stylistic modeling, representative tweets are iteratively sampled across the user timeline and compressed into a concise natural-language style description together with exemplar posts.

These extracted signals collectively provide stable identity priors and temporally grounded behavioral cues for the TWICE generation framework.

\section{Experimental Protocol and Settings}
\label{app:experiment_settings}
\subsection{Event Sampling Protocol}

To ensure temporal representativeness, we adopt a time-weighted event sampling strategy. Events are first grouped by their posting dates, and the relative frequency of events in each temporal segment is used as the sampling weight. User–event instances are then sampled proportionally to these weights, ensuring that both early and recent periods of user activity are represented in the evaluation set.

\subsection{Experimental Settings}

The dataset spans multiple stages of user activity, including disorder onset, progression, and large-scale societal events (e.g., the COVID-19 pandemic). To capture such long-term temporal dynamics, we fix the following parameters unless otherwise specified: retrieval \texttt{time window} = one year, \texttt{node num} = 3, \texttt{memory num} = {10, 20}, temporal decay rate $\lambda = 0.001$, and \texttt{state coeff} = {1, 1.1}. The impact of these parameters is analyzed in RQ2.

\subsection{Baselines}

We compare TWICE against several representative tweet-generation baselines adapted for user simulation:

\begin{itemize}
\item \textbf{Non-personalized}: conditioned only on the current event.
\item \textbf{Memory-only}: uses event-driven memory retrieval without user profiling.
\item \textbf{Profile-only}: conditions on a structured user profile without memory retrieval.
\item \textbf{Implicit profile (USP-style)}: uses a latent persona inferred from long-term tweet history.
\item \textbf{PopAlign persona}: uses a population-aligned narrative persona.
\item \textbf{OASIS}: a social-media agent conditioned on lightweight persona signals and recent tweets.
\end{itemize}

TWICE is evaluated in two variants: \textbf{TWICE (general)} and \textbf{TWICE (event)}, corresponding to general-memory and event-centered retrieval.

\subsection{Temporal Sensitivity Analysis}

To analyze the influence of temporal modeling, we study three key parameters: retrieval \texttt{time window}, \texttt{memory num}, and \texttt{state coeff}.

We also compare four retrieval scoring strategies for selecting top-$K$ memories: (i) similarity-only, (ii) recency-only, (iii) similarity with time decay, and (iv) the full TWICE scoring function, which additionally incorporates memory importance and state-aware weighting. For fair comparison, all variants retrieve the same number of memories and use identical LLM and decoding settings.

\section{Style Consistency Details}
\label{app:style-consistency}

We instantiate the three components of style similarity used in Eq.~\ref{eq:sim-style} as follows.
Given two texts with TF--IDF vectors $\vec{t}_1, \vec{t}_2$ and POS-tag distribution vectors $\vec{p}_1, \vec{p}_2$,
we compute cosine similarities for lexical and syntactic matching:
\begin{subequations}
\label{eq:sim-components}
\begin{align}
\mathrm{Sim}_{\mathrm{tfidf}}
&= \frac{\vec{t}_1 \cdot \vec{t}_2}{\lVert \vec{t}_1 \rVert \, \lVert \vec{t}_2 \rVert}, \label{eq:sim-tfidf}\\
\mathrm{Sim}_{\mathrm{pos}}
&= \frac{\vec{p}_1 \cdot \vec{p}_2}{\lVert \vec{p}_1 \rVert \, \lVert \vec{p}_2 \rVert}. \label{eq:sim-pos}
\end{align}
\end{subequations}

To account for pacing and structural rhythm, we also consider sentence-length statistics.
Let $\mu$ and $\sigma$ denote the mean and standard deviation of sentence lengths (in tokens),
we define
\begin{equation}
\mathrm{Sim}_{\mathrm{length}}
= \frac{1}{1 + \lvert \mu_1 - \mu_2 \rvert + \lvert \sigma_1 - \sigma_2 \rvert}.
\label{eq:sim-length}
\end{equation}

The overall style similarity is defined as:
\begin{equation}
\mathrm{Sim}{\mathrm{style}}
= \frac{\mathrm{Sim}{\mathrm{tfidf}} + \mathrm{Sim}{\mathrm{pos}} + \mathrm{Sim}{\mathrm{length}}}{3}.
\label{eq:sim-style}
\end{equation}

\section{Humanlikeness Details}
\label{app:humanlikeness}

We evaluate \textbf{Humanlikeness} from two complementary aspects: \textit{readability} and \textit{affective complexity}. Given the original tweet $x$ and the simulated tweet $\hat{x}$, we compute:

\paragraph{Readability.}
We use Flesch Reading Ease (FRE) and Flesch--Kincaid Grade Level (FKGL), computed as:
\begin{subequations}
\label{eq:readability}
\begin{align}
\mathrm{FRE} &= 206.835 - 1.015\,\mathrm{ASL} - 84.6\,\mathrm{ASW}, \\
\mathrm{FKGL} &= 0.39\,\mathrm{ASL} + 11.8\,\mathrm{ASW} - 15.59,
\end{align}
\end{subequations}
where $\mathrm{ASL}$ is the average sentence length (words per sentence) and $\mathrm{ASW}$ is the average syllables per word.
We report readability as \textbf{absolute differences} between original and simulated tweets:
\begin{equation}
\begin{aligned}
|\Delta|\mathrm{FRE} &= \left|\mathrm{FRE}(x)-\mathrm{FRE}(\hat{x})\right|,\\
|\Delta|\mathrm{FKGL} &= \left|\mathrm{FKGL}(x)-\mathrm{FKGL}(\hat{x})\right|.
\end{aligned}
\end{equation}

Lower values indicate closer readability to human-written ground-truth tweets.

\paragraph{Affective complexity.}
To quantify affective expression, we compute the KL divergence between valence--arousal--dominance (VAD) distributions extracted from tweets. 
Let $\mathbf{z}_x \in \mathbb{R}^3$ and $\mathbf{z}_{\hat{x}} \in \mathbb{R}^3$ denote the VAD scores of the original and simulated tweets, obtained using a standard affective lexicon model. We convert them into distributions via softmax:
\begin{equation}
P=\operatorname{softmax}(\mathbf{z}_x),\quad Q=\operatorname{softmax}(\mathbf{z}_{\hat{x}}).
\end{equation}
The affective divergence is then:
\begin{equation}
\mathrm{KL}(P\|Q)=\sum_{i=1}^{3} p_i \ln\!\frac{p_i}{q_i}.
\end{equation}
Lower KL divergence indicates the simulated tweet exhibits a more human-like affective structure aligned with the original.

\section{Human Evaluation Details}
\label{appendix:human_eval}

To further validate the realism of generated tweets beyond automatic metrics, we conduct a human evaluation study on the NEG-cohort generalization setting.

\paragraph{Evaluation Setup.}
We randomly sample 50 user--event instances from the shared evaluation pool. For each instance, all compared methods generate outputs conditioned on the same target event to ensure fair comparison. Annotators are shown:
(1) the user's recent tweet history,
(2) retrieved memory tweets,
(3) the target event description,
and (4) anonymized candidate generations presented in randomized order.

We compare three representative approaches:
(1) a non-personalized LLM baseline,
(2) an implicit-profile conditioning baseline following USP-style personalization,
and (3) \textsc{TWICE}.

\paragraph{Annotation Dimensions.}
Annotators rank the generated outputs along three dimensions:

\begin{itemize}[leftmargin=*]
    \item \textbf{Consistency}: whether the generated tweet matches the user's historical linguistic style, tone, and behavioral tendencies;
    
    \item \textbf{Authenticity}: whether the generated content appears plausible and appropriate given both the target event and the user's identity;
    
    \item \textbf{Humanlikeness}: whether the generated tweet resembles a realistic human-authored social media post overall.
\end{itemize}

Lower mean rank indicates better performance.

\paragraph{Annotators and Agreement.}
Three annotators participated in the study, yielding a total of 450 ranked judgments (50 instances $\times$ 3 dimensions $\times$ 3 annotators).

Inter-annotator agreement shows moderate but reliable consistency for this open-ended generation task. As summarized in Table~\ref{tab:human_agreement}, the mean pairwise Kendall's $\tau$ reaches 0.401, indicating stable relative preferences across annotators.

\begin{table}[h]
\centering
\small
\begin{tabular}{lc}
\toprule
Agreement Metric & Value \\
\midrule
All-3 top-1 agreement & 39.3\% \\
Pairwise top-1 agreement & 59.3\% \\
Mean Kendall's $\tau$ & 0.401 \\
\bottomrule
\end{tabular}
\caption{Inter-annotator agreement statistics for human evaluation.}
\label{tab:human_agreement}
\end{table}

\paragraph{Overall Human Preference.}
Table~\ref{tab:human_overall} reports the aggregated ranking results across all dimensions. \textsc{TWICE} achieves the best overall human preference, obtaining both the lowest mean rank and the highest top-1 preference rate.

\begin{table}[h]
\centering
\small
\begin{tabular}{lcc}
\toprule
Method & Mean Rank $\downarrow$ & Top-1 Rate $\uparrow$ \\
\midrule
Non-personalized & 2.16 & 26.9\% \\
Implicit Profile & 1.99 & 31.1\% \\
\textbf{TWICE (ours)} & \textbf{1.85} & \textbf{42.0\%} \\
\bottomrule
\end{tabular}
\caption{Overall human evaluation results aggregated across all dimensions. Lower mean rank and higher top-1 rate indicate better human preference.}
\label{tab:human_overall}
\end{table}

\paragraph{Per-Dimension Analysis.}
Table~\ref{tab:human_dim_appendix} further breaks down performance across the three evaluation dimensions. \textsc{TWICE} consistently achieves the best mean rank in consistency, authenticity, and humanlikeness, suggesting that temporally grounded event conditioning improves not only stylistic fidelity but also perceived behavioral realism.

\begin{table}[h]
\centering
\small
\setlength{\tabcolsep}{3pt}
\begin{tabular}{lccc}
\toprule
Method & Consist. & Auth. & Human. \\
\midrule
Non-personalized & 2.28 & 2.14 & 2.05 \\
Implicit Profile & 1.93 & 1.99 & 2.05 \\
\textbf{TWICE (ours)} & \textbf{1.79} & \textbf{1.87} & \textbf{1.90} \\
\bottomrule
\end{tabular}
\caption{Per-dimension mean ranks in human evaluation (lower is better). 
Consist., Auth., and Human. denote consistency, authenticity, and humanlikeness, respectively.}
\label{tab:human_dim_appendix}
\end{table}

\paragraph{Pairwise Preference Analysis.}
To further analyze comparative preferences, Table~\ref{tab:human_pairwise} reports pairwise win rates of \textsc{TWICE} against the two baselines. \textsc{TWICE} is consistently preferred across all dimensions, with particularly strong gains in stylistic consistency.

\begin{table}[h]
\centering
\small
\setlength{\tabcolsep}{3pt}
\begin{tabular}{lccc}
\toprule
Comparison & Consist. & Auth. & Human. \\
\midrule
TWICE vs Implicit & 57.3 & 54.7 & 56.0 \\
TWICE vs Non-personalized & 64.0 & 58.7 & 54.0 \\
\bottomrule
\end{tabular}
\caption{Pairwise human preference win rates (\%) of \textsc{TWICE} against baselines. 
Consist., Auth., and Human. denote consistency, authenticity, and humanlikeness, respectively.}
\label{tab:human_pairwise}
\end{table}

Overall, the human evaluation results support the findings from automatic metrics: temporally grounded event-driven personalization substantially improves perceived realism, behavioral consistency, and user authenticity.

\section{Retrieval Scoring Variants for Memory Selection}
\label{app:retrieval_scoring}

To analyze how the retrieval strategy affects temporal coherence, we conduct a retrieval ablation by comparing several scoring variants for selecting top-$K$ memory entries. 
All variants share the same retrieval budget ($K$), backbone LLM, prompts, and decoding settings; only the scoring function differs.

\paragraph{Notation.}
Given the current event/context $e_t$ and a candidate memory item $m$, we denote their semantic similarity as $\mathrm{sim}(m,e_t)$ (cosine similarity in embedding space). 
We define $\Delta t$ as the time gap between $m$ and $e_t$, and use a time-decay function:
\begin{equation}
f_{\text{time}}(\Delta t)=\exp(-\lambda \Delta t),
\label{eq:time_decay}
\end{equation}
where $\lambda$ controls the decay rate. 
We further define an importance score $f_{\text{importance}}(m)$ and a state-aware weighting term $f_{\text{state}}(m)$, which increases the retrieval probability for memories that contain time-sensitive state information.

\paragraph{Scoring Variants.}
We consider four scoring schemes:

\begin{enumerate}[leftmargin=1.2em]
    \item \textbf{Similarity-only} ranks memories solely based on semantic relevance:
    \begin{equation}
    s_{\text{sim}}(m \mid e_t)=\mathrm{sim}(m,e_t).
    \label{eq:sim_only}
    \end{equation}

    \item \textbf{Recency-only} prioritizes temporally recent memories:
    \begin{equation}
    s_{\text{rec}}(m \mid e_t)=f_{\text{time}}(\Delta t).
    \label{eq:recency_only}
    \end{equation}

    \item \textbf{Sim$\times$Time-decay} balances topical relevance and temporal locality:
    \begin{equation}
    s_{\text{sim-time}}(m \mid e_t)=\mathrm{sim}(m,e_t)\cdot f_{\text{time}}(\Delta t).
    \label{eq:sim_time}
    \end{equation}

    \item \textbf{Full scoring (\textsc{TWICE})} further incorporates importance and state-aware weighting:
    \begin{equation}
    \begin{aligned}
    s_{\text{full}}(m \mid e_t)
    &= \mathrm{sim}(m,e_t)\cdot f_{\text{time}}(\Delta t) \\
    &\quad \cdot f_{\text{importance}}(m)\cdot f_{\text{state}}(m).
    \end{aligned}
    \label{eq:full_scoring}
    \end{equation}

\end{enumerate}

\paragraph{Protocol.}
For each scoring variant, we retrieve the top-$K$ memories $M_t^{(K)}$ and generate tweets using the same backbone LLM under identical decoding settings. 
We then evaluate the generated outputs using semantic similarity (Sim), readability deviation (FKGL), and length deviation (token difference) relative to the ground-truth posts. 

\begin{table}[t]
\centering
\small
\setlength{\tabcolsep}{4pt}
\begin{tabular}{lccc}
\toprule
\textbf{Retrieval Scoring} & \textbf{Sim} $\uparrow$ & $|\Delta|$ \textbf{FKGL} $\downarrow$ & $\Delta$ \textbf{Tokens} \\
\midrule
Sim-only & 0.7205 & 0.9612 & +13.23 \\
Recency-only & 0.7134 & 3.7623 & +18.03 \\
Sim$\times$Time & 0.7249 & 3.8063 & +14.00 \\
Full scoring (\textsc{TWICE}) & 0.7305 & 0.5653 & +12.31 \\
\bottomrule
\end{tabular}
\caption{Retrieval strategy ablation comparing different memory scoring variants for selecting top-$K$ memories. All variants use the same backbone LLM, decoding settings, and retrieval budget.}
\label{tab:retrieval_ablation_full}
\end{table}

\paragraph{Discussion.}
The results show that similarity-only retrieval yields strong topical alignment but may retrieve generic memories that do not preserve temporal evolution. 
Recency-only improves locality but can miss semantically critical historical events. 
In contrast, combining similarity with time decay provides a stronger balance between topical relevance and temporal coherence. 
Finally, the full \textsc{TWICE} scoring achieves the best overall trade-off, suggesting that integrating importance and state-sensitive factors is beneficial for faithful temporally-grounded simulation.

\section{Event Detection and Structured Extraction}
\label{app:event_extraction}

To construct event-driven memory nodes, TWICE adopts a two-stage pipeline for identifying and structuring life events from user timelines.

First, tweets are filtered using supervised classifiers from PsyEvent~\cite{Chen2024MappingLC} and PsySym~\cite{Zhang2022SymptomIF}. These models assign relevance scores for different life-event categories and behavioral signals. Tweets whose scores exceed a predefined confidence threshold are retained as candidate event posts.

Second, a large language model performs structured event extraction on the filtered tweets. The extraction process identifies the main event triple together with associated metadata fields, including the event type, emotional signal, time expressions, location references, and the user's role in the event. The extracted information is normalized into RDF-style triples to support downstream memory construction.

Finally, extracted events are grouped and summarized into compact event memories, which serve as temporally grounded context for the TWICE simulation framework.

The detailed prompts used for event extraction are provided in Appendix~\ref{app:prompt_event}.

\section{Prompt}
\label{app:prompt_event}
In this section, we provide the details of the prompts used for different stages of the user simulator. These prompts include extracting demographic information from the user's posts, identifying significant events from their timeline, and simulating the posting workflow. The prompts also include event detection and timeline extraction to simulate posts based on recent events.

\subsection{Extracting Profile}
The \textbf{Extracting Profile} prompt is designed to extract demographic information from the user's posts, including details like age, occupation, and other personal characteristics that may be reflected in their posts. 

\begin{tcolorbox}[colback=gray!10, 
                 colframe=black, 
                 width=\linewidth, 
                 arc=2mm, 
                 title={Infer Age Prompt}, breakable]
    Infer the age(int) of the user until now(2021.1.1). \\
    "timestamp\_tweet" is the time when the user posted this tweet, "text" is the content of the user's tweets. \\
    If you cannot infer the age according to the tweets, the answer will be: None. \\
    Attention: if there is enough information to infer the age, you need to calculate the user's age based on the current date (2021.1.1). For example, if the user's tweet in 2013 stated 'I'm 21', his age should be: 21 + (2020 -2013) = 28. \\
    The output of the age must be in the form of an integer (int). \\
    Today is 2021.1.1. Here are the tweets that a user posted on the social media: \\
    !\{tweets\}! \\
    Please strictly output a JSON object with the following format, no additional text: \\
    \{ \\
        "age": int, \\
        "explanation": "your explanation" \\
    \} \\
    ONLY output valid JSON, WITHOUT any extra characters or explanations.
\end{tcolorbox}

\begin{tcolorbox}[colback=gray!10, 
                 colframe=black, 
                 width=\linewidth, 
                 arc=2mm, 
                 title={Personality Analysis Prompt}, breakable]
    You are an expert in computational psychology and linguistic analysis. \\
    Please analyze the user's personality based on their tweets below, according to the !\{dimension\}! trait: \\
    !\{tweets\}! \\
    Requirements: \\
    1. Provide a qualitative rating of Low, Medium, or High based on the language patterns, emotional tone, and content in the user's tweets. \\
    2. Provide a detailed explanation (1-3 sentences) justifying the rating for the !\{dimension\}! trait. \\
    3. Focus exclusively on the user’s characteristics, disregarding any information related to others, unless it directly impacts the user. \\
    4. !\{dimension\}! trait definition: \\
    !\{definition\}! \\
    5. Please strictly output a JSON object with the following format, no additional text: \\
    \{ \\
        "score": "Low/Medium/High", \\
        "explanation": "Explanation referencing specific tweet evidence." \\
    \} \\
    ONLY output valid JSON, WITHOUT any extra characters or explanations.
\end{tcolorbox}

\begin{tcolorbox}[colback=gray!10, 
                 colframe=black, 
                 width=\linewidth, 
                 arc=2mm, 
                 title={Infer Marital Status Prompt}, breakable]
    Infer the latest marital status(married/divorced/single/widowed/unknown) of the user until now(2021.1.1). \\
    "timestamp\_tweet" is the time when the user posted this tweet, "text" is the content of the user's tweets. \\
    Here are some explanation for the classification of the marital status: \\
    \textbf{married}: the user has married with his/her spouse still alive without divorce. \\
    \textbf{divorced}: the user has married with his/her spouse still alive, but they divorced. \\
    \textbf{single}: the user has never married. \\
    \textbf{widowed}: the user has married and hasn't divorced, but his/her spouse died. \\
    If you cannot infer the marital status according to the tweets, the answer will be: None. \\
    Attention: if there is enough information to infer the marital status, you have to choose one status from married/divorced/single/widowed/unknown. You can't make up other description. Your answer must be strictly based on the tweets and your explanation should be brief. \\
    Today is 2021.1.1. Here are the tweets that a user posted on the social media: \\
    !\{tweets\}! \\
    Please strictly output a JSON object with the following format, no additional text: \\
    \{ \\
        "marital\_status": "married/divorced/single/widowed/unknown", \\
        "explanation": "your explanation" \\
    \} \\
    ONLY output valid JSON, WITHOUT any extra characters or explanations.
\end{tcolorbox}

\begin{tcolorbox}[colback=gray!10, 
                 colframe=black, 
                 width=\linewidth, 
                 arc=2mm, 
                 title={Infer Work Status Prompt}, breakable]
    Infer the latest work status of the user\\
    (employed / unemployed / retired /\\
    student / unknown) until now (2021.1.1). \\
    "timestamp\_tweet" is the time when the user posted this tweet, "text" is the content of the user's tweets. \\
    If you cannot infer the work status according to the tweets, the answer will be: None. \\
    Attention: if there is enough information to infer the work status, you have to choose one status from employed/unemployed/retired/student/unknown. You can't make up other description. Your answer must be strictly based on the tweets and your explanation should be brief. \\
    Today is 2021.1.1. Here are the tweets that a user posted on the social media: \\
    !\{tweets\}! \\
    Please strictly output a JSON object with the following format, no additional text: \\
    \{ \\
        "work\_status": "employed/unemployed/retired/student/unknown", \\
        "explanation": "your explanation" \\
    \} \\
    ONLY output valid JSON, WITHOUT any extra characters or explanations.
\end{tcolorbox}

\begin{tcolorbox}[colback=gray!10, 
                 colframe=black, 
                 width=\linewidth, 
                 arc=2mm, 
                 title={Infer Career Domain Prompt}, breakable]
    Here is the self-description of a twitter user: \\
    !\{description\}! \\
    
    Based on the description, please infer the career domain of the user. \\
    If there is enough information, you have to choose the career domain strictly from the nine domains: \\
    Creative Arts and Media (represented by 0), Business and Finance (represented by 1), Technology and Engineering (represented by 2), Healthcare and Social Services (represented by 3), Education and Research (represented by 4), Legal and Public Policy (represented by 5), Transportation and Logistics (represented by 6), Manufacturing and Construction (represented by 7), Hospitality and Tourism (represented by 8). \\
    
    Here are the descriptions for the nine domains above: \\
    0. Creative Arts and Media: Focus on artistic expression and media production, including visual arts, film, music, design, and journalism, where creativity and communication are key. \\
    1. Business and Finance: Centered around managing financial assets, business operations, marketing, investment, and sales, driving economic growth and organizational success. \\
    2. Technology and Engineering: Involve applying scientific principles to design and develop technology, software, infrastructure, and solutions to solve complex problems. \\
    3. Healthcare and Social Services: Focus on providing medical care, mental health support, and social welfare services to improve the well-being of individuals and communities. \\
    4. Education and Research: Dedicate to teaching, academic research, and fostering learning and development through formal and informal education systems. \\
    5. Legal and Public Policy: Concern with law enforcement, legal counsel, and shaping public policy to ensure justice, governance, and the protection of rights. \\
    6. Transportation and Logistics: Focus on the efficient movement of goods and people, including transportation planning, logistics management, and supply chain operations. \\
    7. Manufacturing and Construction: Relate to the production of physical goods, construction of buildings and infrastructure, and managing the process of creating tangible products or structures. \\
    8. Hospitality and Tourism: Focus on providing services related to travel, accommodation, and leisure, including hotels, restaurants, tour operations, and event planning. \\
    
    Please strictly output a JSON object with the following format, no additional text: \\
    \{ \\
        "career\_domain": int(0/1/2/3/4/5/6/7/8), \\
        "explanation": "your explanation" \\
    \} \\
    ONLY output valid JSON, WITHOUT any extra characters or explanations. \\
    Your inference must be strictly based on the description, and you should only output the corresponding field number if relevant terminology related to the field is explicitly mentioned in the description. Unwarranted assumptions and extrapolations are not allowed. \\
    If you cannot infer the career domain according to the description, the output will be: None.
\end{tcolorbox}

\begin{tcolorbox}[colback=gray!10, 
                 colframe=black, 
                 width=\linewidth, 
                 arc=2mm, 
                 title={Analyze Posting Style Prompt}, breakable]
    !\{posts\}! \\
    
    Analyze the above Twitter posts from a user and summarize their posting style. Please consider the following aspects in your analysis: \\
    1. \textbf{Tone and Sentiment}: Determine the general tone (e.g., positive, negative, neutral) and emotional intensity (e.g., calm, passionate, sarcastic). \\
    2. \textbf{Language Complexity}: Assess the complexity of the language used in the posts. Are the sentences short and direct, or are they more complex with multiple clauses? \\
    3. \textbf{Vocabulary and Word Choice}: Identify the kind of vocabulary the user employs. Is the language formal or informal? Does the user use industry-specific jargon, slang, or creative expressions (e.g., abbreviations, memes)? \\
    4. \textbf{Topic Focus}: What are the common themes or topics discussed in the posts? Are they related to politics, technology, entertainment, personal life, or any specific interest areas? \\
    5. \textbf{Engagement and Interactions}: How does the user interact with others? Are there frequent replies, retweets, mentions, or hashtags? How does the user engage with other Twitter users in their responses? \\
    6. \textbf{Cultural or Social Context}: Does the user reference any cultural, social, or political issues? How do these references influence their tone or language? \\
    7. \textbf{Creativity and Innovation}: Does the user employ creative language or innovative expressions (e.g., playful language, spelling alterations, new phrases)? \\
    8. \textbf{Post Length and Structure}: Are the posts long and detailed or short and concise? How does the user structure their tweets (e.g., paragraphs, bullet points, or single sentences)? \\
    
    Based on these aspects, describe the user's overall posting style in a concise and fluent paragraph, no more than 100 words. \\
    
    Please strictly output a JSON object with the following format, no additional text: \\
    \{ \\
        "description": "your description" \\
    \} \\
    ONLY output valid JSON, WITHOUT any extra characters or explanations.
\end{tcolorbox}

\begin{tcolorbox}[colback=gray!10, 
                 colframe=black, 
                 width=\linewidth, 
                 arc=2mm, 
                 title={Select 20 Best Tweets Prompt}, breakable]
    Here are some tweets of a user, please select the 20 tweets that you believe best reflect the user's expressive style and language habits: \\
    !\{tweets\}! \\
    
    Your answer should consist of two parts: the IDs of the 20 tweets that best reflect the user's expressive style and language habits, and your reasons and criteria for your selection. \\
    
    Please strictly output a JSON object with the following format, no additional text: \\
    \{ \\
        "tweet\_id": [id\_1(int), id\_2(int), ..., id\_20(int)], \\
        "explanation": "your explanation" \\
    \} \\
    ONLY output valid JSON, WITHOUT any extra characters or explanations.
\end{tcolorbox}

\subsection{Extracting Event}
The \textbf{Extracting Event} prompt is used to identify significant events from the user's posts that may influence their emotional state or behavior. These events are crucial for simulating a realistic reaction from the user in response to specific events. Additionally, the prompt identifies events that may be linked or form an event cluster. The goal is to create event-driven simulations of user behavior.

\begin{tcolorbox}[colback=gray!10, 
                 colframe=black, 
                 width=\linewidth, 
                 arc=2mm, 
                 title={Event Information Extraction Prompt}, breakable]
    You are a social media event information extraction expert. Here is a tweet about !\{item\}! : \\
    !\{tweet\}! \\
    
    First, you need to determine whether the event related to the !\{item\}! mentioned in this tweet is meaningful to the user. \\
    If not, your output will be: None. \\
    
    If the event related to the !\{item\}! mentioned in this tweet is meaningful to the user, extract the main RDF-style event triple about !\{item\}! and provide structured metadata fields. \\
    This will be used as an external stimulus for generating user responses later. \\
    
    Return output in JSON format, with the following fields: \\
    
    \begin{itemize}
        \item \texttt{event\_triple}: The main RDF-style triple, formatted as \textless subject\textgreater \textless predicate\textgreater \textless object\textgreater
        \item \texttt{event\_type}: One of ['Career', 'Death', 'Education', 'Financial', 'Health', 'Identity', 'Legal', 'Lifestyle\_Change', 'New\_Birth\_in\_Family', 'Relationships\_Changes', 'Relocation', 'Societal']
        \item \texttt{emotion}: One of ["Joy", "Trust", "Fear", "Surprise", "Sadness", "Disgust", "Anger", "Anticipation", "Neutral", "Mixed"].
        \item \texttt{time\_expression}: Any time-related phrase (e.g., “yesterday”), or null.
        \item \texttt{location\_expression}: Any location-related phrase (e.g., “parking garage”), or null.
        \item \texttt{external\_events}: If the tweet references or implies an external public event, social issue, or controversy, describe it concisely; otherwise, return null.
        \item \texttt{related\_context}: Optional. If there is any personal, social, identity-related, or historical background that helps explain why the event occurred, include a brief explanation; otherwise, return null.
        \item \texttt{surface\_variants}: Provide different sentence variants to express the same event.
        \item \texttt{user\_role}: Specify the role of the user in the event. Possible values: ["initiator", "experiencer", "observer", "object"].
    \end{itemize}
    
    Rules:
    \begin{itemize}
        \item Extract only objective, real-world events explicitly or clearly implied in the post. These events will be used to simulate user responses to external stimuli.
        \item The subject should reflect the real-world actor, which may be the "User", another person, company, society, etc.
        \item Do NOT assume the subject is always the User; the subject should be the real agent involved in the event.
        \item Use normalized, canonical phrases for predicates.
        \item Do NOT hallucinate; only extract what is explicitly or reasonably implied.
        \item If possible, include time or location as an auxiliary triple.
    \end{itemize}

    Examples:
    \begin{itemize}
        \item Employee requested remote work due to COVID concerns, but was terminated by the company.
    \end{itemize}

    Please strictly output a JSON object with the following format, no additional text: \\
    \{ \\
        "event\_triple": "\textless subject\textgreater \textless predicate\textgreater \textless object\textgreater", \\
        "event\_type": "", \\
        "emotion": "", \\
        "time\_expression": "", \\
        "location\_expression": "", \\
        "external\_events": "", \\
        "related\_context": "", \\
        "surface\_variants": [""], \\
        "user\_role": "" \\
    \} \\
    ONLY output valid JSON, WITHOUT any extra characters or explanations.
\end{tcolorbox}

\begin{tcolorbox}[colback=gray!10, 
                 colframe=black, 
                 width=\linewidth, 
                 arc=2mm, 
                 title={Event Relation Identification Prompt}, breakable]
    Here are some tweets related to !\{event\}! Your task is to determine whether any of these tweets describe events that are strongly related to each other, either by forming a temporal sequence (i.e., a timeline) or by representing similar types of events occurring at different times. The events must be directly experienced by the user — do not include posts that only describe someone else’s experience. \\
    !\{tweets\}! \\
    
    Requirements: \\
    1. The events you identify must be clearly related to !\{event\}! and must be directly experienced by the user — do not include posts that only describe someone else’s experience. \\
    2. The tweets must not be repeated expressions of the same event on the same day — you can refer to the timestamp\_tweet to ensure the tweets are from different days. \\
    3. Your goal is to identify clusters or sequences of related events, and briefly explain the relationship between them. \\
    4. Your output should include three parts: the tweet\_id of the tweets you find; the brief conclusion of the event that these tweets related to; your explanation. \\
    5. Please strictly output a JSON object with the following format, no additional text: \\
    \{ \\
        "tweet\_id": [tweet\_id\_1(int), tweet\_id\_2(int), ..., tweet\_id\_n(int)], \\
        "event\_conclusion": "", \\
        "explanation": "" \\
    \} \\
    ONLY output valid JSON, WITHOUT any extra characters or explanations. \\
    6. If there aren't any tweets that are highly related to each other and refer to similar events, your response should be: \\
    \{ \\
        "tweet\_id": None, \\
        "event\_conclusion": None, \\
        "explanation": None \\
    \}
\end{tcolorbox}

\subsection{Simulating Posting Workflow}
The \textbf{Simulating Posting Workflow} involves generating tweets style rewriting. 

\begin{tcolorbox}[colback=gray!10, 
                 colframe=black, 
                 width=\linewidth, 
                 arc=2mm, 
                 title={Simulated Tweet Generation Prompt}, breakable]
    You are a twitter user. \\
    This is your profile: \\
    !\{profile\}! \\
    Now something has happened to you: \\
    !\{event\}! \\
    Here are your previous posts ("timestamp\_tweet" is the time when you posted the tweet): \\
    !\{memory\}! \\
    
    Based on your memory and profile, imagine your reaction towards this event, and write a tweet. \\
    Your should imitate the language style of your previous tweets and try your best to act like the user. \\
    Your response should not contradict your previous tweets or your profile. \\
    Your response should be a tweet composed of one or a few sentences. \\
    
    Attention: You should try your best to imitate the language style and tone based on previous tweets! \\
    You can imitate the tone of the user: \\
    !\{style\_tweets\}! \\
    
    Please strictly output a JSON object with the following format, no additional text: \\
    \{ \\
        "simulated\_tweet": "your simulated tweet" \\
    \} \\
    ONLY output valid JSON, WITHOUT any extra characters or explanations.
\end{tcolorbox}

\begin{tcolorbox}[colback=gray!10, 
                 colframe=black, 
                 width=\linewidth, 
                 arc=2mm, 
                 title={Rewriting Prompt}, breakable]
    You are an expert in analyzing and mimicking social media users' posting styles. \\
    Please rewrite the following tweet to match the user's style. Use the provided user's Big Five personality traits, some of their past tweets, and a summary of their posting style to guide your rewrite. \\
    
    User's Big Five Personality Traits: \\
    !\{big\_five\}! \\
    
    Original Tweet to Rewrite: \\
    !\{simulated\_tweet\}! \\
    
    !\{style\}! \\
    
    Requirements: \\
    1. Maintain the same general content and message as the original tweet. \\
    2. Modify the tone, word choice, structure, and phrasing to reflect the user's unique personality and posting style (if you know). \\
    3. Make sure the rewrite feels authentic to the user, as if it could have been written by them. \\
    4. The rewritten tweet should match the user's personality traits, past posting history (if you know), and overall style summary (if you know). \\
    
    Please strictly output a JSON object with the following format, no additional text: \\
    \{ \\
        "rewritten\_tweet": " ", \\
        "explanation": "your explanation and how you imitate the user's style" \\
    \} \\
    ONLY output valid JSON, WITHOUT any extra characters or explanations.
\end{tcolorbox}

\section{Case Study}
In this section, we show the whole procedure of our framework.

\subsection{Original Data}
For each user, the original data include two parts: account information and tweets. In the following two text boxes, we provide the user's account information and the user's original posts published on the same day.

\begin{tcolorbox}[colback=gray!10, 
                 colframe=black, 
                 width=\linewidth, 
                 arc=2mm, 
                 title={Account info}, breakable]
    \textbf{Creation Timestamp:} 2010-08-31 12:04:37+00:00 \\
    \textbf{Description:} bdh fan | i like watching movies and tv shows so i decided to work on them! \\
    \textbf{Favourites Count:} 6673 \\
    \textbf{Followers Count:} 2659 \\
    \textbf{Friends Count:} 184 \\
    \textbf{User ID:} 185178713 \\
    \textbf{Geo Tag:} (none) \\
    \textbf{Status Count:} 130714 \\
    \textbf{Verified Check:} No
\end{tcolorbox}

\begin{tcolorbox}[colback=gray!10, 
                 colframe=black, 
                 width=\linewidth, 
                 arc=2mm, 
                 title={Tweet 1}, breakable]
\textbf{Tweet ID:} 1398726440703741957 \\
\textbf{Timestamp:} 2021-05-29 19:42:25+00:00 \\
\textbf{Language:} English \\
\textbf{Source:} Twitter for iPhone \\
\textbf{Likes Count:} 1 \\
\textbf{Quote Count:} 0 \\
\textbf{Reply Count:} 1 \\
\textbf{Retweet Count:} 0 \\
\textbf{Mentioned Users:} 2232006592 \\
\textbf{Text:} Yes but maybe they decided to change things a little bit? Maybe we won’t have an official trailer until the end of summer/later this year but they might give us a little something for the original released date! Fingers crossed
\end{tcolorbox}

\subsection{Profile}
Through techniques such as large model inference, regularization matching, semantic similarity calculation, and fine-tuned BERT models as life event and symptom classifiers, we extract the user's profile from the raw data. A specific profile is shown as follows:

\begin{tcolorbox}[colback=gray!10, 
                 colframe=black, 
                 width=\linewidth, 
                 arc=2mm, 
                 title={User Profile}, breakable]
\textbf{User ID:} 185178713 \\
\textbf{Age:} 26 \\
\textbf{Gender:} Female \\
\textbf{Marital Status:} Single \\
\textbf{Career Domain:} Creative Arts and Media \\
\textbf{Work Status:} Employed \\
\textbf{Description:} bdh fan | i like watching movies and tv shows so i decided to work on them!  \\
\textbf{Creation Timestamp:} 2010-08-31 12:04:37+00:00 \\
\textbf{Favourites Count:} 6673 \\
\textbf{Followers Count:} 2659 \\
\textbf{Friends Count:} 184 \\
\textbf{Geo Tag:} (none) \\
\textbf{Status Count:} 130714 \\
\textbf{Verified Check:} No \\
\textbf{Big Five Personality Traits:} \\
\hspace{1em}\textbf{Openness:} Medium \\
\hspace{1em}\textbf{Conscientiousness:} Medium \\
\hspace{1em}\textbf{Extraversion:} Medium \\
\hspace{1em}\textbf{Neuroticism:} Medium \\
\hspace{1em}\textbf{Agreeableness:} Medium \\
\textbf{Life Events:} \\
\hspace{1em}\textbf{Career:} Experience varied career challenges and opportunities in the film industry, including job transitions, early starts, and balancing multiple projects. \\
\hspace{1em}\textbf{Death:} Experience death through laughter over a friend's comment about fashion, mourning a cousin's death described as an angel joining the sky, grieving over a pet cat's passing, reacting to a young student's fatal drug overdose with disbelief, and mentioning an emotional impact from an unspecified reason related to Bryce Dallas Howard. \\
\hspace{1em}\textbf{Education:} Graduate from film school with mixed feelings, face challenges like thesis stress, online class anxiety, and job uncertainties post-graduation. \\
\hspace{1em}\textbf{Financial:} Express concern over high healthcare costs and insurance coverage issues in the USA, noting extreme cases like exorbitant bills for childbirth and lack of coverage for certain services. \\
\hspace{1em}\textbf{Health:} Experience various health issues including physical pain, mental health challenges, and medical appointments while navigating healthcare systems and treatments. \\
\hspace{1em}\textbf{Identity:} Expresses pride in women speaking out against abuse while also doubting the effectiveness of such actions in getting abusers to come forward. \\
\hspace{1em}\textbf{Legal:} Experience legal threats ranging from parking disputes to copyright infringement notices, involving potential police involvement and family separation due to traffic violations related to immigration enforcement. \\
\hspace{1em}\textbf{Lifestyle Change:} Strive for healthier lifestyles including fitness, better eating habits, and mental well-being, while also facing challenges like maintaining motivation and dealing with personal stresses. \\
\hspace{1em}\textbf{New Birth in Family:} Expecting and welcoming new family members, including a baby cousin and an adopted pet, with joy and anticipation, while also experiencing the loss of a baby cousin. \\
\hspace{1em}\textbf{Relationships Changes:} Express excitement for others' relationship developments and mention personal struggles with crushes and hurt feelings. \\
\hspace{1em}\textbf{Relocation:} Plan trips and experience mixed emotions including excitement and anxiety about relocating to different places such as New Zealand, Australia, Mexico, Scotland, and the USA. \\
\hspace{1em}\textbf{Societal:} Experience societal events including lockdowns, voting frustrations, climate change concerns, and reactions to global incidents like the Paris attacks and natural disasters. \\
\hspace{1em}\textbf{Anxious Mood:} Experience frequent anxiety triggered by various situations including new jobs, family health issues, social interactions, and personal milestones, often accompanied by feelings of stress, fear, and physical discomfort. \\
\hspace{1em}\textbf{Autonomic Symptoms:} Experience headaches and rapid heartbeat, sometimes attributing symptoms to lack of sleep or overthinking. \\
\hspace{1em}\textbf{Cardiovascular Symptoms:} Experience cardiovascular symptoms including heart palpitations and chest pain, with tweets expressing distress and seeking help. \\
\hspace{1em}\textbf{Catatonic Behavior:} (none) \\
\hspace{1em}\textbf{Decreased Energy/Tiredness/Fatigue:} Feel exhausted frequently, experiencing fatigue that affects daily life and sleep patterns. \\
\hspace{1em}\textbf{Depressed Mood:} Feel frequent sadness, crying, and hopelessness due to depression and anxiety, often exacerbated by external triggers and internal struggles with self-esteem and familial support. \\
\hspace{1em}\textbf{Gastrointestinal Symptoms:} Feel nausea and gastrointestinal discomfort, including a sensation of something burning and stuck in the throat. \\
\hspace{1em}\textbf{Genitourinary Symptoms:} (none) \\
\hspace{1em}\textbf{Hyperactivity/Agitation:} (none) \\
\hspace{1em}\textbf{Impulsivity:} Wait for conditions or events related to personal relationships and health issues, expressing frustration and desire for change regarding these situations. \\
\hspace{1em}\textbf{Inattention:} (none) \\
\hspace{1em}\textbf{Indecisiveness:} (none) \\
\hspace{1em}\textbf{Respiratory Symptoms:} Hate coughing and experiencing chest pain. \\
\hspace{1em}\textbf{Suicidal Ideas:} Express suicidal thoughts and feelings of despair across multiple occasions. \\
\hspace{1em}\textbf{Worthlessness and Guilty:} Feel worthlessness and guilty due to persistent self-esteem issues, depression, and anxiety, exacerbated by perceived failures and societal comparisons. \\
\hspace{1em}\textbf{Avoidance of Stimuli:} (none) \\
\hspace{1em}\textbf{Compensatory Behaviors to Prevent Weight Gain:} (none) \\
\hspace{1em}\textbf{Compulsions:} (none) \\
\hspace{1em}\textbf{Diminished Emotional Expression:} Feel reliving the Bataclan attack emotionally. \\
\hspace{1em}\textbf{Do Things Easily Get Painful Consequences:} Experience headaches easily after drinking little, indicating age-related changes. \\
\hspace{1em}\textbf{Drastical Shift in Mood and Energy:} Express mood swings and difficulties waking up in a good mood due to factors like dreams, Sunday nights, and general energy levels. \\
\hspace{1em}\textbf{Fear About Social Situations:} Struggle with social anxiety affecting everyday life including education, employment, and interpersonal relationships, while feeling unsupported by parents. \\
\hspace{1em}\textbf{Fear of Gaining Weight:} (none) \\
\hspace{1em}\textbf{Fears of Being Negatively Evaluated:} Let fears of being negatively evaluated cause hurt and financial stress, as seeking help like therapy is hindered by the cost and societal pressures related to mental health. \\
\hspace{1em}\textbf{Flight of Ideas:} (none) \\
\hspace{1em}\textbf{Intrusion Symptoms:} Overthink, stress, and struggle with insomnia. \\
\hspace{1em}\textbf{Loss of Interest or Motivation:} Lose motivation for everything due to recent events, mentioning specific struggles with starting an internship. \\
\hspace{1em}\textbf{More Talkative:} (none) \\
\hspace{1em}\textbf{Obsession:} Expresses an obsession with movies and movie making. \\
\hspace{1em}\textbf{Panic Fear:} Expresses feeling stressed to the point of joking about nearly dying from it. \\
\hspace{1em}\textbf{Pessimism:} (none) \\
\hspace{1em}\textbf{Poor Memory:} Feel distressed by poor memory, experiencing nightmares and emotional distress when trying to recall details. \\
\hspace{1em}\textbf{Sleep Disturbance:} Struggle to fall asleep due to various factors including excitement, anxiety, overthinking, and physical discomfort, leading to poor sleep quality and short sleep duration. \\
\hspace{1em}\textbf{Somatic Muscle:} Experience pain in the neck, back, and shoulders, seeking chiropractic help. \\
\hspace{1em}\textbf{Somatic Symptoms (Others):} Experience frequent crying spells and physical discomfort due to anxiety. \\
\hspace{1em}\textbf{Somatic Symptoms (Sensory):} Experience extreme sensitivity to temperature, expressing discomfort in both hot and cold conditions. \\
\hspace{1em}\textbf{Weight and Appetite Change:} (none) \\
\hspace{1em}\textbf{Anger/Irritability:} Expresses frequent irritation and anger towards family and coworkers, struggling with perceptions shaped by these emotions.
\end{tcolorbox}

\subsection{Event Tweets}
Through LLM inference, we have identified and extracted posts containing significant events from the user, along with posts from the timeline related to those events. The following two text boxes display a post containing an important event and a timeline of the user's related posts:

\begin{tcolorbox}[colback=gray!10, 
                 colframe=black, 
                 width=\linewidth, 
                 arc=2mm, 
                 title={Event Tweet: First Appointment with Therapist}, breakable]
\textbf{Tweet ID:} 1285264241784758274 \\
\textbf{Timestamp:} 2020-07-20 17:24:08+00:00 \\
\textbf{Language:} English \\
\textbf{Source:} Twitter for iPhone \\
\textbf{Likes Count:} 2 \\
\textbf{Quote Count:} 0 \\
\textbf{Reply Count:} 0 \\
\textbf{Retweet Count:} 0 \\
\textbf{Mentioned Users:} None \\
\textbf{Media:} None \\
\textbf{Life Event:} Health \\
\textbf{Text:} i had my first appointment with my therapist today.. i’m glad i finally went even though i was apprehensive about it! \\
\textbf{Event Conclusion:} Had a first appointment with a therapist. \\
\textbf{Event Extract:} \\
\hspace{1em} \textbf{Event Triple:} \textless User\textgreater \textless attended first appointment with\textgreater \textless therapist\textgreater \\
\hspace{1em} \textbf{Event Type:} Health \\
\hspace{1em} \textbf{Emotion:} Mixed \\
\hspace{1em} \textbf{Time Expression:} Today \\
\hspace{1em} \textbf{Location Expression:} None \\
\hspace{1em} \textbf{External Events:} None \\
\hspace{1em} \textbf{Related Context:} User had a first therapy session after feeling apprehensive about it. \\
\hspace{1em} \textbf{Surface Variants:} \\
\hspace{2em} 1. I attended my first therapy session today despite initial hesitation. \\
\hspace{2em} 2. Today, I went to my first appointment with a therapist. \\
\hspace{2em} 3. Despite my apprehensions, I went to see my therapist for the first time today. \\
\hspace{1em} \textbf{User Role:} Experiencer
\end{tcolorbox}

\begin{tcolorbox}[colback=gray!10, 
                 colframe=black, 
                 width=\linewidth, 
                 arc=2mm, 
                 title={Tweet 1: First Appointment with Therapist}, breakable]
\textbf{Tweet ID:} 1285264241784758274 \\
\textbf{Timestamp:} 2020-07-20 17:24:08+00:00 \\
\textbf{Language:} English \\
\textbf{Source:} Twitter for iPhone \\
\textbf{Likes Count:} 2 \\
\textbf{Quote Count:} 0 \\
\textbf{Reply Count:} 0 \\
\textbf{Retweet Count:} 0 \\
\textbf{Text:} i had my first appointment with my therapist today.. i’m glad i finally went even though i was apprehensive about it! \\
\textbf{Event:} <User> <attended first appointment with> <therapist> \\
\textbf{Emotion:} Mixed \\
\textbf{Time Expression:} Today \\
\textbf{Related Context:} User had a first therapy session after feeling apprehensive about it. \\
\textbf{Surface Variants:} \\
\hspace{1em} 1. I attended my first therapy session today despite initial hesitation. \\
\hspace{1em} 2. Today, I went to my first appointment with a therapist. \\
\hspace{1em} 3. Despite my apprehensions, I went to see my therapist for the first time today. \\
\textbf{User Role:} Experiencer
\end{tcolorbox}

\vspace{1em}

\begin{tcolorbox}[colback=gray!10, 
                 colframe=black, 
                 width=\linewidth, 
                 arc=2mm, 
                 title={Tweet 2: Taking Appointment with Therapist}, breakable]
\textbf{Tweet ID:} 1283722480364990465 \\
\textbf{Timestamp:} 2020-07-16 11:17:44+00:00 \\
\textbf{Language:} English \\
\textbf{Source:} Twitter for iPhone \\
\textbf{Likes Count:} 0 \\
\textbf{Quote Count:} 0 \\
\textbf{Reply Count:} 1 \\
\textbf{Retweet Count:} 0 \\
\textbf{Text:} I took an appointment with a therapist. I’ve been postponing doing it for the past two years. I’m terrified  \\
\textbf{Event:} <User> <took an appointment with> <therapist> \\
\textbf{Emotion:} Fear \\
\textbf{Time Expression:} Two years \\
\textbf{Related Context:} User has been postponing the appointment for two years and feels terrified. \\
\textbf{Surface Variants:} \\
\hspace{1em} 1. I finally made an appointment with a therapist after postponing for two years. \\
\hspace{1em} 2. After two years of procrastination, I scheduled a therapy session. \\
\hspace{1em} 3. I'm seeing a therapist today, something I've put off for two long years. \\
\textbf{User Role:} Initiator
\end{tcolorbox}

\vspace{1em}

\begin{tcolorbox}[colback=gray!10, 
                 colframe=black, 
                 width=\linewidth, 
                 arc=2mm, 
                 title={Tweet 3: Planning to Find a Therapist}, breakable]
\textbf{Tweet ID:} 1211373980080386048 \\
\textbf{Timestamp:} 2019-12-29 19:50:37+00:00 \\
\textbf{Language:} English \\
\textbf{Source:} Twitter for iPhone \\
\textbf{Likes Count:} 2 \\
\textbf{Quote Count:} 0 \\
\textbf{Reply Count:} 0 \\
\textbf{Retweet Count:} 0 \\
\textbf{Text:} First thing I’ll do in 2020 is find a therapist because I just can’t anymore \\
\textbf{Event:} <User> <plans to seek> <therapy> \\
\textbf{Emotion:} Sadness \\
\textbf{Time Expression:} In 2020 \\
\textbf{Related Context:} The user expresses feeling overwhelmed and unable to cope without professional help. \\
\textbf{Surface Variants:} \\
\hspace{1em} 1. In 2020, I plan to find a therapist because I just can't anymore. \\
\hspace{1em} 2. Next year, I'm going to look for a therapist since I've reached my limit. \\
\hspace{1em} 3. I've decided to see a therapist at the start of 2020 due to my current struggles. \\
\textbf{User Role:} Initiator
\end{tcolorbox}

\vspace{1em}

\begin{tcolorbox}[colback=gray!10, 
                 colframe=black, 
                 width=\linewidth, 
                 arc=2mm, 
                 title={Tweet 4: Struggles with Seeking Therapy}, breakable]
\textbf{Tweet ID:} 1050471916648185856 \\
\textbf{Timestamp:} 2018-10-11 19:43:16+00:00 \\
\textbf{Language:} English \\
\textbf{Source:} Twitter for iPhone \\
\textbf{Likes Count:} 0 \\
\textbf{Quote Count:} 0 \\
\textbf{Reply Count:} 1 \\
\textbf{Retweet Count:} 0 \\
\textbf{Text:} I’m so sorry :( I 100
\textbf{Event:} <User> <needs to seek> <therapy> \\
\textbf{Emotion:} Fear \\
\textbf{Time Expression:} Two years ago \\
\textbf{Related Context:} User acknowledges the need for therapy but expresses fear and reluctance about seeking it again. \\
\textbf{Surface Variants:} \\
\hspace{1em} 1. I need to see a therapist again but I'm scared. \\
\hspace{1em} 2. It's been two years since I last saw a therapist and I know I need to start again. \\
\hspace{1em} 3. Seeing a therapist helped me before, but now I'm afraid to take that first step. \\
\textbf{User Role:} Experiencer
\end{tcolorbox}

\subsection{Memory}
We designed two types of memory: general memory and event memory. General memory is aggregated by time, where posts from the same month form a memory node. Event memory is aggregated based on events as features, where posts related to the same event form a memory node. The following two text boxes show examples of general memory nodes and event memory nodes. In this system, the \textit{Tweet Importance} is initialized to 1 and will be incremented during the retrieval process based on the retrieval results. The \textit{Tweet Embedding} is obtained using Qwen3-Embedding-8B, and it is a multidimensional list. Due to its length, it is not displayed here.

\begin{tcolorbox}[colback=gray!10, 
                 colframe=black, 
                 width=\linewidth, 
                 arc=2mm, 
                 title={General Memory Node for December 2020}, breakable]
\textbf{Node Time:} 2020-12-30 \\
\textbf{Number of Tweets:} 626 \\
\textbf{Node Importance:} 1 \\
\textbf{Node Embedding:} 1 \\
\textbf{Tweets:}

\begin{itemize}
    \item \textbf{Tweet ID:} 1344173588597911558 \\
          \textbf{Timestamp:} 2020-12-30 06:48:51+00:00 \\
          \textbf{Text:} LMFAO \#MyTwitterAnniversary  \\
          \textbf{Tweet Importance:} 1 \\
          \textbf{Tweet Embedding:} 1

    \item \textbf{Tweet ID:} 1343978086157721602 \\
          \textbf{Timestamp:} 2020-12-29 17:52:00+00:00 \\
          \textbf{Text:} To the toilet? \\
          \textbf{Tweet Importance:} 1 \\
          \textbf{Tweet Embedding:} 1

    \item \textbf{Tweet ID:} 1343977587371089921 \\
          \textbf{Timestamp:} 2020-12-29 17:50:01+00:00 \\
          \textbf{Text:} I like how y'all are gathering them  \\
          \textbf{Tweet Importance:} 1 \\
          \textbf{Tweet Embedding:} 1
\end{itemize}

\end{tcolorbox}

\begin{tcolorbox}[colback=gray!10, 
                 colframe=black, 
                 width=\linewidth, 
                 arc=2mm, 
                 title={Event Memory for Career - December 2020}, breakable]
\textbf{Node Time:} 2020-12-11 \\
\textbf{Number of Tweets:} 94 \\
\textbf{Node Importance:} 1 \\
\textbf{Node Embedding:} 1 \\
\textbf{Tweets:}

\begin{itemize}
    \item \textbf{Tweet ID:} 1337201855869366273 \\
          \textbf{Timestamp:} 2020-12-11 01:05:41+00:00 \\
          \textbf{Text:} I like how every employed person I know has decided to rock regardless of the day \\
          \textbf{Event:} Career \\
          \textbf{Tweet Importance:} 1 \\
          \textbf{Tweet Embedding:} 1

    \item \textbf{Tweet ID:} 1336469265860485121 \\
          \textbf{Timestamp:} 2020-12-09 00:34:38+00:00 \\
          \textbf{Text:} Looking forward to getting a decent job (or rich husband) so I can let people enjoy their money  \\
          \textbf{Event:} Career \\
          \textbf{Tweet Importance:} 1 \\
          \textbf{Tweet Embedding:} 1

    \item \textbf{Tweet ID:} 1334416714998624256 \\
          \textbf{Timestamp:} 2020-12-03 08:38:31+00:00 \\
          \textbf{Text:} Someone here isn't focusing on their job \\
          \textbf{Event:} Career \\
          \textbf{Tweet Importance:} 1 \\
          \textbf{Tweet Embedding:} 1
\end{itemize}

\end{tcolorbox}

\subsection{Simulation Result}
We first perform event extraction on a post that contains significant events experienced by the user, and the extracted events are summarized as sudden events, with the original post's posting time recorded as the time of the occurrence of the event. All posts made by the user prior to the event are considered as the memory available to the user simulator, and we use retrieval techniques to select the top-k most relevant memories. The event summary, retrieved memories, and the user profile are provided to a large model for the first simulation generation. After obtaining the first generation result, we perform style rewriting to make the simulated post's style more human-like and better reflect the user's personalized characteristics. In the second rewriting, we provide the first simulation result, the user's Big Five personality traits, and the user's posting style (including a summary of the user's posting style and representative posts reflecting their style) to the large model for further rewriting. The second rewritten result typically shows better performance in terms of authenticity, consistency, and humanlikeness.
The following text boxes provide the user's original post, the extracted event, the user's event profile, the retrieved memories, the first generation result, the user's style description, the user's Big Five personality traits, and the second rewritten result.

\begin{tcolorbox}[colback=gray!10, 
                 colframe=black, 
                 width=\linewidth, 
                 arc=2mm, 
                 title={Original User's Tweet}, breakable]
\textbf{Tweet ID:} 1200231490409373698 \\
\textbf{Timestamp:} 2019-11-29 01:54:21+00:00 \\
\textbf{Text:} Update: I went to the doctor because of the slump I was going through and ended up diagnosed with a wombo combo of anxiety and severe depression on top of my already diagnosed ADHD. How I managed to graduate like this is a mystery.  \\
\textbf{Likes:} 3 \\
\textbf{Replies:} 1 \\
\textbf{Retweets:} 0 \\
\textbf{Source:} Twitter for Android
\end{tcolorbox}

\begin{tcolorbox}[colback=gray!10, 
                 colframe=black, 
                 width=\linewidth, 
                 arc=2mm, 
                 title={Extracted Event: Health - Severe Depression}, breakable]
\textbf{Event Triple:} \\
\textless User\textgreater \textless was diagnosed with\textgreater \textless severe depression\textgreater \\
\textbf{Event Type:} Health \\
\textbf{Time Expression:} recently \\
\textbf{Location Expression:} null \\
\textbf{External Events:} null \\
\textbf{Related Context:} User sought medical attention due to a persistent slump and received multiple diagnoses. \\
\textbf{Surface Variants:}
\begin{itemize}
    \item I recently got diagnosed with severe depression.
    \item After visiting the doctor, I found out I have severe depression.
    \item Recently diagnosed with severe depression while feeling down.
\end{itemize}
\textbf{User Role:} experiencer
\end{tcolorbox}

\begin{tcolorbox}[colback=gray!10, 
                 colframe=black, 
                 width=\linewidth, 
                 arc=2mm, 
                 title={User's Event Profile}, breakable]
\textbf{User ID:} 72986462 \\
\textbf{Age:} 27 \\
\textbf{Gender:} Female \\
\textbf{Marital Status:} Single \\
\textbf{Career Domain:} Creative arts and media \\
\textbf{Work Status:} Employed \\
\textbf{Description:} Illustrator and concept artist || I love everything DnD and cyberpunk stuff. Also trying to make games in my spare time. She/they \\
\textbf{Creation Timestamp:} 2009-09-09 23:48:05+00:00 \\
\textbf{Followers Count:} 250 \\
\textbf{Friends Count:} 1062 \\
\textbf{Status Count:} 4135 \\
\textbf{Likes Count:} 10549 \\
\textbf{Geo-tag:} Probably in front of my pc \\
\textbf{Verified Check:} No \\
\textbf{Event Profile:} \\
\textbf{Career:} Experience varied career challenges and opportunities in the film industry, including job transitions, early starts, and balancing multiple projects. \\
\textbf{Death:} Lose oldest friend nine months before pandemic, father dies from cancer/pneumonia in August, struggle with second birthday of friend who passed, reflect on six months since father's death, express ongoing regrets over limited time with father. \\
\textbf{Education:} Experience ADHD undiagnosed in education system, struggle with thesis and finals, and balance creative pursuits like animation and concept art with academic pressures. \\
\textbf{Financial:} Manage financial aspects like pricing, payments, and budgeting while facing unexpected expenses such as veterinary costs for rescued animals. \\
\textbf{Health:} Experiencing various health issues including depression, anxiety, ADHD, and physical ailments like gastritis and anemia, while navigating treatment and its effects on daily life and mental health. \\
\textbf{Identity:} Explore identity and sexuality through various personal and artistic expressions, including coming to terms with being gay, creating lesbian characters, and navigating disclosure with family. \\
\textbf{Lifestyle Change:} Consider quitting art, planning new hairstyles, adjusting routine, and managing anxiety while continuing creative endeavors like painting and streaming. \\
\textbf{Relationships Changes:} Experience changes in relationships, including conflicts with a best friend, emotional detachment leading to cutting off contact, and reflections on past friendships and attractions. \\
\textbf{Relocation:} Camps in the living room for five months due to remodeling and expects to stay for two more weeks until fully settled. \\
\textbf{Societal:} Support Peruvian Lives Matter and protest against government actions, gaining attention on Twitter.
\end{tcolorbox}

\begin{tcolorbox}[colback=gray!10, 
                 colframe=black, 
                 width=\linewidth, 
                 arc=2mm, 
                 title={First Simulation Generation Result}, breakable]
After visiting the doctor, I found out I have severe depression. Guess I'm just another heart patient with a side of existential dread.
\end{tcolorbox}

\begin{tcolorbox}[colback=gray!10, 
                 colframe=black, 
                 width=\linewidth, 
                 arc=2mm, 
                 title={Retrieved Event Memory: Health-Related Tweets}, breakable]
\textbf{Node Name:} Death \\
\textbf{Final Score:} 0.0329 \\
\textbf{Similarity:} 0.2394 \\
\textbf{Golden Similarity:} 0.1587 \\
\textbf{Time Weight:} 0.1249 \\
\textbf{Importance Weight:} 1.1 \\
\textbf{Tweet:} 
\begin{itemize}
    \item \textbf{Text:} *tells best friend lady mormont actress voices hilda\\
          \textbf{Timestamp:} 2019-05-04 23:25:46+00:00 \\
          \textbf{Tweet ID:} 1124817424833044480
\end{itemize}

\vspace{1em}

\textbf{Node Name:} Cardiovascular Symptoms \\
\textbf{Final Score:} 0.0031 \\
\textbf{Similarity:} 0.5045 \\
\textbf{Golden Similarity:} 0.3874 \\
\textbf{Time Weight:} 0.0056 \\
\textbf{Importance Weight:} 1.1 \\
\textbf{Tweet:} 
\begin{itemize}
    \item \textbf{Text:} My heart can't handle this  \\
          \textbf{Timestamp:} 2018-06-27 15:20:22+00:00 \\
          \textbf{Tweet ID:} 1011992643675574273
\end{itemize}

\vspace{1em}

\textbf{Node Name:} Weight and Appetite Change \\
\textbf{Final Score:} 0.0026 \\
\textbf{Similarity:} 0.3467 \\
\textbf{Golden Similarity:} 0.2993 \\
\textbf{Time Weight:} 0.0069 \\
\textbf{Importance Weight:} 1.1 \\
\textbf{Tweet:} 
\begin{itemize}
    \item \textbf{Text:} I'm eating out with some friends today but they are used to eat a lot later than me so now I'm starving \\
          \textbf{Timestamp:} 2018-07-18 18:13:26+00:00 \\
          \textbf{Tweet ID:} 1019646341725802496
\end{itemize}

\vspace{1em}

\textbf{Node Name:} Cardiovascular Symptoms \\
\textbf{Final Score:} 0.0024 \\
\textbf{Similarity:} 0.5045 \\
\textbf{Golden Similarity:} 0.3874 \\
\textbf{Time Weight:} 0.0044 \\
\textbf{Importance Weight:} 1.1 \\
\textbf{Tweet:} 
\begin{itemize}
    \item \textbf{Text:} My heart can't handle this  \\
          \textbf{Timestamp:} 2018-06-03 02:32:51+00:00 \\
          \textbf{Tweet ID:} 1003102183053840384
\end{itemize}

\vspace{1em}

\textbf{Node Name:} Cardiovascular Symptoms \\
\textbf{Final Score:} 0.0019 \\
\textbf{Similarity:} 0.4262 \\
\textbf{Golden Similarity:} 0.3273 \\
\textbf{Time Weight:} 0.0041 \\
\textbf{Importance Weight:} 1.1 \\
\textbf{Tweet:} 
\begin{itemize}
    \item \textbf{Text:} This is why I get my heart stolen every 2.5 seconds  \\
          \textbf{Timestamp:} 2018-05-27 10:08:55+00:00 \\
          \textbf{Tweet ID:} 1000680241096994816
\end{itemize}

\vspace{1em}

\textbf{Node Name:} Respiratory Symptoms \\
\textbf{Final Score:} 0.0017 \\
\textbf{Similarity:} 0.3995 \\
\textbf{Golden Similarity:} 0.384 \\
\textbf{Time Weight:} 0.0038 \\
\textbf{Importance Weight:} 1.1 \\
\textbf{Tweet:} 
\begin{itemize}
    \item \textbf{Text:} Me: I think I'm starting to get sick\\
          *proceeds to ignore symptoms for the next 3 days\\
          Now guess who has trouble breathing \\
          \textbf{Timestamp:} 2018-05-22 01:22:53+00:00 \\
          \textbf{Tweet ID:} 998735917698486272
\end{itemize}

\vspace{1em}

\textbf{Node Name:} Do Things Easily Get Painful Consequences \\
\textbf{Final Score:} 0.0017 \\
\textbf{Similarity:} 0.3115 \\
\textbf{Golden Similarity:} 0.3883 \\
\textbf{Time Weight:} 0.0048 \\
\textbf{Importance Weight:} 1.1 \\
\textbf{Tweet:} 
\begin{itemize}
    \item \textbf{Text:} Tbh I don't want her to go if she's not ready, she should actually open up to mina before that. Going can make things worse since her anxiety is going to build up, it can bring all your insecurities back in a sec and get you all self destructive. Face your fears but do so wisely. \\
          \textbf{Timestamp:} 2018-06-14 01:22:21+00:00 \\
          \textbf{Tweet ID:} 1007070704913829888
\end{itemize}

\vspace{1em}

\textbf{Node Name:} Compensatory Behaviors to Prevent Weight Gain \\
\textbf{Final Score:} 0.0011 \\
\textbf{Similarity:} 0.3618 \\
\textbf{Golden Similarity:} 0.3849 \\
\textbf{Time Weight:} 0.0027 \\
\textbf{Importance Weight:} 1.1 \\
\textbf{Tweet:} 
\begin{itemize}
    \item \textbf{Text:} I'm not sure how is it even possible to get distracted this fast. \\
          \textbf{Timestamp:} 2018-04-16 01:25:17+00:00 \\
          \textbf{Tweet ID:} 985690558369845249
\end{itemize}

\end{tcolorbox}

\begin{tcolorbox}[colback=gray!10, 
                 colframe=black, 
                 width=\linewidth, 
                 arc=2mm, 
                 title={User's Style Characteristics}, breakable]
\textbf{Style Partial:} \\
The user posts with a mix of tones, often sarcastic or emotionally intense. Language is informal with slang, abbreviations, and creative expressions. Topics range widely but lean towards entertainment, personal life, and fandoms. Posts are brief and engaging, referencing memes and cultural trends without frequent interactions beyond personal expressions.

\textbf{Style Overall:} \\
The user's tweets exhibit a mix of personal reflections and engagement with popular culture, particularly focusing on gaming, anime, and K-pop. The tone varies from casual and humorous to passionate and politically charged, reflecting both personal feelings and broader social issues. Language is informal with frequent use of slang, emojis, and creative expressions. Posts often include hashtags and interactions with others, showing active engagement.
\end{tcolorbox}

\begin{tcolorbox}[colback=gray!10, 
                 colframe=black, 
                 width=\linewidth, 
                 arc=2mm, 
                 title={User's Style Tweets}, breakable]
\textbf{Tweet 1:} \\
LMFAO \#MyTwitterAnniversary 

\textbf{Tweet 2:} \\
Yo, I’m just living for the chaos. 

\textbf{Tweet 3:} \\
INTERRUMPIMOS LA PROGRAMACION PARA MORIR POR K/DA. SO GOOD, MARRY ME EVE. 

\textbf{Tweet 4:} \\
grey's anatomy mode, 4 temporadas en 2 semanas(tengo problemas)

\textbf{Tweet 5:} \\
NOOOOOOOOOOOOOOOOO!!! fin del mundo, bram!, noooo, todo mal todomal todokfjaknascdas. i need my brittana NOT bram!!! *tears* T-T *again*
\end{tcolorbox}

\begin{tcolorbox}[colback=gray!10, 
                 colframe=black, 
                 width=\linewidth, 
                 arc=2mm, 
                 title={Second Rewriting Result}, breakable]
DOC SAID I'M A HEART PATIENT WITH EXISTENTIAL DREAD  PTM ALERT! \#GreyAnatomyMode \#KDAForever
\end{tcolorbox}

\end{document}